\documentclass[lettersize,journal]{IEEEtran}
\usepackage{cite,graphicx,psfig,amsmath,amssymb,eufrak,mathrsfs,epsf,epsfig}
\usepackage{psfrag}
\usepackage{caption}
\usepackage{subcaption}
\usepackage{array}
\usepackage{multirow}
\usepackage{algpseudocode,algorithmicx}
\usepackage{algorithm}
\usepackage{verbatim}
\usepackage{epstopdf}
\usepackage[x11names]{xcolor}

\makeatletter
\newcommand{\SplitState}[1]{%
  \State
  \parbox[t]{\dimexpr\linewidth-\ALG@thistlm}{%
    #1\par\xdef\Split@prevdepth{\the\prevdepth}%
  }\par
  \prevdepth\Split@prevdepth
}
\makeatother

\usepackage[active]{srcltx}
\ifCLASSINFOpdf

\else
\fi

\begin{document}
\newtheorem{ach}{Achievability}
\newtheorem{con}{Converse}
\newtheorem{definition}{Definition}
\newtheorem{theorem}{Theorem}
\newtheorem{lemma}{Lemma}
\newtheorem{example}{Example}
\newtheorem{cor}{Corollary}
\newtheorem{prop}{Proposition}
\newtheorem{conjecture}{Conjecture}
\newtheorem{remark}{Remark}
\title{Hybrid Online--Offline Learning for Task Offloading in Mobile Edge Computing Systems}
\author{\IEEEauthorblockN{Muhammad Sohaib, Sang-Woon Jeon,~\IEEEmembership{Senior Member,~IEEE}, and Wei Yu,~\IEEEmembership{Fellow,~IEEE}}


\thanks{The work of M. Sohaib and S.-W. Jeon was supported by the Korea Evaluation Institute Of Industrial Technology (KEIT) grant funded by the Korean Government [Korean Coast Guard (KCG), Ministry of the Interior and Safety (MOIS), and National Fire Agency (NFA)] (Development of technology to respond to marine fires and chemical accidents using wearable devices) under Grant RS-2022-001549812. The work of W. Yu was supported by the Natural Sciences and Engineering Research Council via the Canada Research Chairs program.}%

\thanks{M. Sohaib is with the Department of Electronics Engineering, Hanyang University, Ansan 15588, South Korea (e-mail: msohaib828@hanyang.ac.kr).}%
\thanks{S.-W. Jeon, corresponding author, is with the School of Computer Science and Technology, Zhejiang
Normal University, Jinhua 321004, China and the Department of Electrical and Electronic Engineering, Hanyang University, Ansan 15588, South Korea (e-mail: sangwoonjeon@hanyang.ac.kr).}%
\thanks{W. Yu is with the Edward S. Rogers Sr. Department of Electrical and Computer Engineering, University of Toronto, Toronto, ON M5S3G4, Canada (e-mail: weiyu@ece.utoronto.ca).}%
}

 \maketitle

\begin{abstract}
We consider a multi-user multi-server mobile edge computing (MEC) system, in which users arrive on a network randomly over time and generate computation tasks, which will be computed either locally on their own computing devices or be offloaded to one of the MEC servers. Under such a dynamic network environment, we propose a novel task offloading policy based on hybrid online--offline learning, which can efficiently reduce the overall computation delay and energy consumption only with information available at nearest MEC servers from each user. We provide a practical signaling and learning framework that can train deep neural networks for both online and offline learning and can adjust its offloading policy based on the queuing status of each MEC server and network dynamics. Numerical results demonstrate that the proposed scheme significantly reduces the average computation delay for a broad class of network environments compared to the conventional offloading methods. It is further shown that the proposed hybrid online--offline learning framework can be extended to a general cost function reflecting both delay- and energy-dependent metrics. 
\end{abstract}
\begin{IEEEkeywords}
Computation offloading, deep learning, mobile edge computing, reinforcement learning, resource allocation.
\end{IEEEkeywords}
 \IEEEpeerreviewmaketitle

\section{Introduction}  \label{intro}
With the explosive growth of intelligent mobile and internet of things (IoT) devices, there has been a tremendous increase in data computation, storage and communication requirements. In order to resolve these massive challenges, cloud computing \cite{cloudsurvey} was initially proposed. Cloud servers generally have a large pool of computation and storage resources that can be robustly deployed and released. But they are physically far from users (end devices) and running all applications on cloud can easily lead to network congestion. To provide ultra-low latency services required by various real-time applications running on end devices, mobile edge computing (MEC) \cite{surveymec} has emerged as a promising technology in efficiently addressing such latency, which can be seen as an extension of cloud computing, but brings computation processing in close proximity to end devices. This proximity reduces computation delay, while also providing higher bandwidth and more flexible computing.

Several works have explored the integration of cloud and edge computing to provide low-latency computing and storage services to end devices \cite{CaoBinLSMODOES,CaoBinRAIoVASFCC,heidari2020internet,HArashDQLTOOCBEGI,XiaozhuMOPTOCCD2D,DaixingxiaTOCAFC}. This integration combines the advantages of both paradigms by delegating computational tasks to edge devices and using the cloud for storage and processing that require more resources. Specifically, a study on the internet of vehicles (IoV) has addressed the placement problem of edge servers, considering multiple objectives such as transmission delay, workload balancing, and energy consumption \cite{CaoBinLSMODOES}. Fog computing and software-defined networking based on fifth-generation (5G) IoV architectures have been studied in \cite{CaoBinRAIoVASFCC}. In the context of IoT, \cite{heidari2020internet} provided an overview of IoT offloading methods and related platforms such as edge and fog computing that are interconnected or dedicated to IoT and \cite{HArashDQLTOOCBEGI} presented a deep Q-learning approach for IoT-edge offloading enabled blockchain. Additionally, joint optimization of content caching and parallel task offloading has been studied in \cite{XiaozhuMOPTOCCD2D, DaixingxiaTOCAFC}. These works contribute to the understanding and advancement of the integration between cloud and edge computing paradigms.

In order to unleash the full potential of MEC systems, however, an intelligent task offloading policy that can efficiently exploit limited resources in edge servers is required \cite{surveytaskoffloading}. 
In \cite{JiangHongboJTORAEMEC}, an online joint offloading and resource allocation framework for MEC networks has been proposed, to ensure quality of experience for end devices. In \cite{DaixingxiaTCOD2DMEC}, a learning-based offloading algorithm for device-to-device (D2D) assisted MEC networks has been studied. In \cite{MitsiouDTAMECGFA}, a novel partial offloading scheme has been proposed to minimize average delay for digital twin aided MEC systems.
In recent studies, deep reinforcement learning (DRL) has been widely adopted to develop optimal task offloading policies for various design objectives \cite{LiDRLoffloadresource, chen_energyoffload,CaoDRLJointaccessoffload}. 
The works in \cite{LiDRLoffloadresource,chen_energyoffload} propose a DRL based energy-efficient task offloading and resource allocation scheme. The work in \cite{CaoDRLJointaccessoffload} proposes a multi-agent DRL scheme to reduce the computation delay and improve the success rate for channel access.
In addition, several DRL-based task offloading schemes have also been developed for MEC systems incorporated with vehicular devices \cite{Livehicular} and unmanned aerial vehicle (UAV) based edge computing units \cite{saccoUavofload,zhaouav} by considering user mobility or optimal UAV positions/trajectories for computation offloading.

Previous works for MEC systems have been mainly focused on the joint optimization of offloading strategies including allocation of communication and computation resources to minimize the computing latency and/or energy consumption under a centralized agent or decentralized multi-agent setup when a set of tasks and resources are given in each time slot~\cite{CaoDRLJointaccessoffload, LiDRLoffloadresource, chen_energyoffload,Livehicular,saccoUavofload,zhaouav}.   
Also, there exist recent works considering random arrival of computation tasks for MEC systems in which computation tasks arrive randomly at each time slot according to some probability distributions~\cite{energydelaysamrat,chen_ptimoffloavirtual,Ming_DRLoffloadqueue}. Specifically, the entire computation resource is divided into several sub-units to serve multiple tasks at each time slot in~\cite{energydelaysamrat} and separate queuing for each user has been considered in~\cite{Ming_DRLoffloadqueue}.

It is worthwhile to mention that the key technical methodology of the previous works for providing intelligent task offloading to MEC systems is to apply reinforcement learning (RL). However, directly applying RL in an online manner may be inefficient due to issues such as signaling overhead, computational complexity, and limitation of online learning.
In this paper, we assume a common queue at each edge server so that a newly arrived task will be served in a first-in first-out (FIFO) manner, which might be suitable for the random arrival model. We further assume a general time-varying computation capability for each edge server that will be affected by several external processes handled by each server. Under such a general framework, we propose a hybrid online--offline task offloading policy that can efficiently minimize the overall delay and energy consumption of arrived tasks. More specifically, the expected delay and energy consumption are predicted by an estimator based on historical information of network dynamics and these predicted values are utilized by an RL agent for offloading decision. The main contributions are summarized as follows:
\begin{itemize}
\item We propose a generalized edge computing framework that can efficiently handle randomly arrived computation tasks in real time under practical signaling between users and their adjacent edge servers by adapting its offloading policy based on local information reported from adjacent edge servers.
\item We propose an online--offline learning methodology for determining a MEC policy of each arrived task. Specifically, we firstly train delay and energy estimators in an offline manner that predicts the expected delay and energy respectively for computing a task at a specific edge server. The trained estimator is then utilized by an RL agent to determine a MEC policy in real time, which is trained in an online manner.
\item Comprehensive simulation results are provided to demonstrate the performance improvement of the proposed scheme compared with the existing MEC strategies. In particular, the proposed scheme can achieve improved performance compared to the conventional RL approaches.
 It is further shown that the proposed hybrid online--offline learning framework is applicable for a broad class of cost functions including both delay- and energy-dependent metrics.
\end{itemize}

\subsection{Related Work}
In recent years, a number of studies have aimed to enhance the capabilities of various types of MEC systems by presenting solutions that reduce computation latency and/or decrease energy consumption\cite{Bozorg_MOCSEDCMECE,GuoYing_DMLMMECS,Zhouwen_DCOMIMOMECSEH,Tuong_PCONOMAMEC,Deng_UCCOEC,AleLaha_DAECOMECDRL,Feng_JTPUALMMEC,zhu_EOMTCDNMEC}. These two fundamental metrics, computation latency and energy consumption, have been widely used to evaluate the performance of MEC systems. To achieve such goals, researchers have developed efficient techniques for allocating communication and computation resources, assigning users to specific MEC servers, determining the optimal proportion of tasks to offload, and etc.
In \cite{Bozorg_MOCSEDCMECE}, the authors proposed an evolutionary computation algorithm based task offloading policy in order to minimize the energy consumption and task processing delay.
In \cite{GuoYing_DMLMMECS}, a federated learning framework for a multi-user MEC system to optimize the task offloading, bandwidth, and computational capability allocation ratio was studied in order to minimize both the energy and delay performances.
In \cite{Zhouwen_DCOMIMOMECSEH}, Lyapunov optimization and successive convex approximation were considered to solve the problem of computation offloading in a multiple-input and multiple-output (MIMO) enabled MEC system in order to minimize the weighted average of latency and energy consumption.
In \cite{Tuong_PCONOMAMEC}, a DRL based algorithm was proposed to reduce the computational overhead (weighted sum of energy and delay) in a non-orthogonal multiple access (NOMA) assisted MEC network by jointly optimizing the computation offloading policy and channel resource allocation under dynamic network environments with time-varying channels. 
In \cite{Deng_UCCOEC}, the authors proposed a particle swarm optimization algorithm to find an optimal loading scheme to minimize the weighted cost of time delay, energy consumption, and price.
In \cite{AleLaha_DAECOMECDRL}, an end-to-end DRL method was studied to select the best edge server for offloading and allocate the optimal computational resource such that the expected long-term utility is maximized, i.e., maximizing the completed tasks before their respective deadlines and at same time minimizing energy consumption.
In addition to the joint minimization of energy and delay, there exist several works focusing on the problem of latency minimization.
In \cite{Feng_JTPUALMMEC}, the authors proposed a dual decomposition-based approach or a matching-based approach to solve the problem of task partitioning and user association in an MEC system in order to minimize the average latency of all users. 
In \cite{zhu_EOMTCDNMEC}, a DRL based method to minimize the task computation delay for a single-user multi-edge server MEC network was proposed. DRL algorithms have been also applied for beam tracking in massive MIMO systems \cite{jeong2020online} and for random-access systems to guarantee fair resource sharing between multiple users in a decentralized manner \cite{Sohaib}. An online learning algorithm for estimating the number of active users in random-access systems has been proposed in \cite{Jeon}.

Although the above mentioned works have proposed various effective algorithms for different types of MEC systems, there are still certain limitations and simplicities present in both the user and MEC models. 
Most previous works considered a joint optimization of different parameters such as selection of an offloaded server, power and bandwidth allocation, and computation resource allocation. Note that such joint optimization can only be performed at a centralized controller with complete information about the entire system, which might be quite challenging in practice due to network dynamics and signaling overhead.
To address these limitations, we assume that users randomly generate tasks over time and the communication and computation resources available at each MEC server are also time-varying. Under such network dynamics, we propose a hybrid online--offline task offloading policy that can efficiently determine an offloading policy of each arrived task in real time with a limited amount of information exchange. 

\begin{figure}[h]
	\centering
	\includegraphics[scale=0.42]{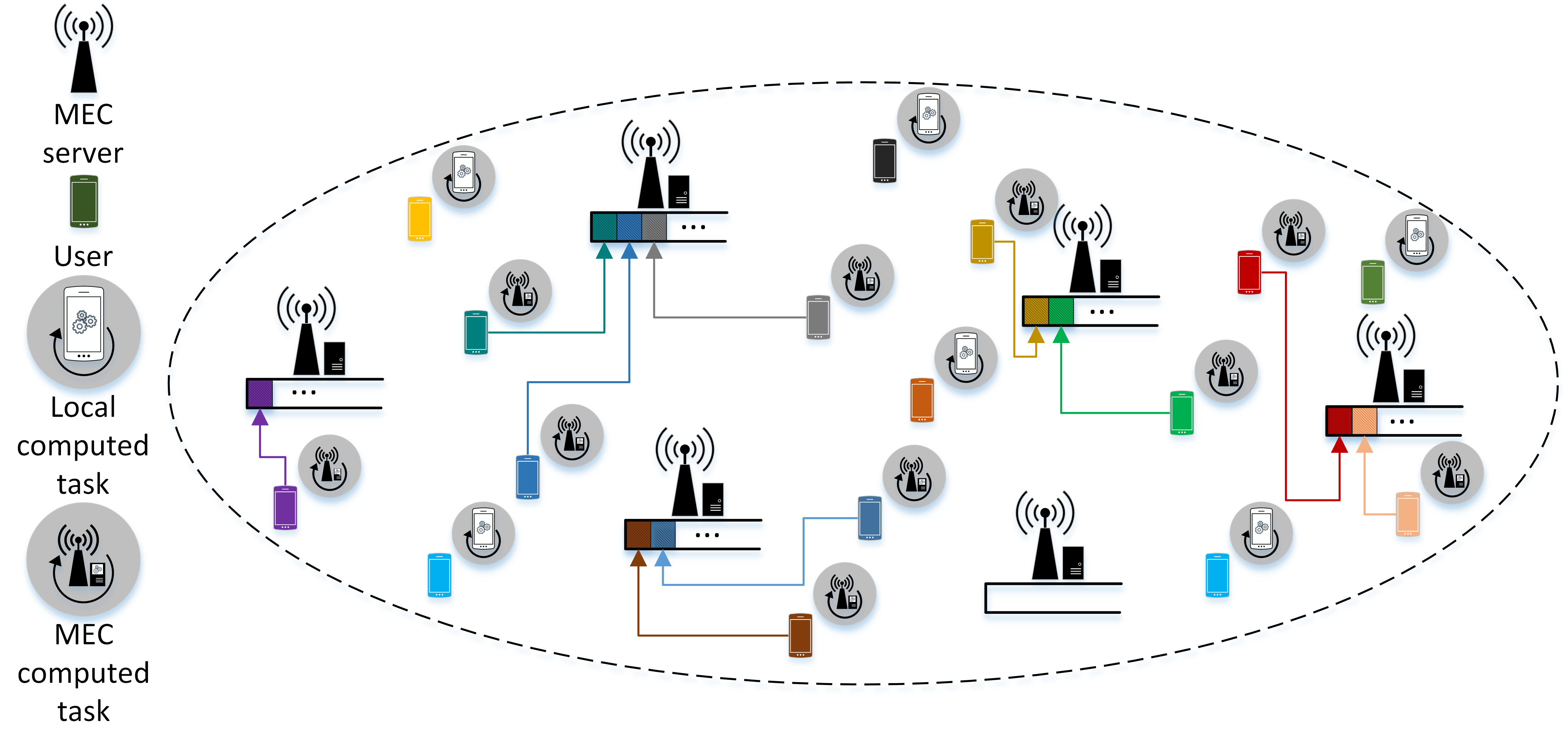}
	\centering
	\caption{Computation task offloading for MEC systems.}
	\label{fig:main_mec_system}
\end{figure}

\subsection{Notation and Paper Organization}
Throughout the paper, $[1:n]$ denotes $\{1,2\cdots, n\}$.
For a vector $\mathbf{a}$, denote the $i$th element of $\mathbf{a}$ by $[\mathbf{a}]_i$.
For a matrix $\mathbf{A}$, denote the $(i,j)$th element of $\mathbf{A}$ by $[\mathbf{A}]_{i,j}$ or $a[i,j]$.
For a set $\mathcal{A}$, denote the cardinality of $\mathcal{A}$ by $|\mathcal{A}|$.
The indicator function is denoted by $\mathbf{1}_{(\cdot)}$. Also $\|\cdot\|$ denotes the norm of a vector and $\mathbf{0}_a$ denotes the all-zero vector of size $a$.

This paper is organized as follows. In Section \ref{sec:PF}, we state the delay minimization of MEC systems considered in this paper. In Section \ref{sec:rl}, we propose a hybrid online--offline learning scheme for computation task offloading. In Section \ref{sec:extension}, we discuss about the extension of the proposed MEC framework to a general cost function including energy consumption for computing. In Section \ref{sec:simulation}, we provide numerical evaluation of the proposed scheme and compare it with the existing MEC strategies. Finally, concluding remarks are given in Section \ref{sec:conclusion}.

\section{Problem Formulation} \label{sec:PF}
In this section, we introduce the system model and the performance metrics used throughout the paper. We focus on the delay minimization in this section and then state how to extend it to a general cost function reflecting both delay- and energy-dependent metrics in Section \ref{sec:extension}.

\subsection{Network Model} \label{Network_Model}
We consider a MEC system as illustrated in Fig. \ref{fig:main_mec_system}, consisting of $M$ distributed edge servers in a network area $[0,a]^2$. 
The position of server $m$ is denoted by  $\mathbf{u}_m\in [0,a]^2$.
We assume that orthogonal bandwidth $W$ [Hz] is assigned to each server so that $N$ independent communication channels are available at each server by dividing $W$ into $N$ sub-bands each of which is allocated to each channel, i.e., frequency division multiple access (FDMA).

We assume that users appear in a network uniformly at random with arrival rate $\lambda$ [the average number of arrivals/sec] and each user generates a computation task or job that can either be computed by its own computing device or be offloaded to one of the servers for computation. 
Specifically, denote the task of user $k$ by $J_k\triangleq(\alpha_k,\beta_k)$ and its arrival time by $t^{[\text{arr}]}_k$ [second], where $\alpha_k$ [bits] is the size of input data required to start the computation task and $\beta_k$ [CPU cycles] is the required amount of computation for completing the task.
Suppose that the position of user $k$ is given by $\mathbf{u}^{[\text{user}]}_k\in[0,a]^2$. Then the channel coefficient from user $k$ to channel $n$ of server $m$ is represented by
\begin{align}
h_k[m,n]=\frac{g_k[m,n]}{\big\|\mathbf{u}_m-\mathbf{u}^{[\text{user}]}_k\big\|^{\gamma/2}}.
\end{align}
Here, $\gamma\geq 2$ denotes the path-loss exponent and $g_k[m,n]$ denotes the small-scaling fading that follows circularly symmetric complex Gaussian distribution with zero mean and unit variance, i.e., $\mathcal{CN}(0,1)$.
Because bandwidth $W/N$ is assigned to each channel, the rate of 
\begin{align} \label{eq:uploading_rate}
r_k[m,n]=\frac{W}{N}\log\left(1+\frac{|h_k[m,n]|^2P_k}{\frac{W}{N}N_0}\right)
\end{align}
is achievable if user $k$ accesses to sever $m$ using channel $n$, where $P_k$ [Watts] is the transmit power of user $k$ and $N_0$ [Watts/Hz] is the noise power density. Denote the $M\times N$ dimensional rate matrix of user $k$ as 
\begin{align}
\mathbf{R}_k=
  \left[ {\begin{array}{ccc}
    r_{k}[1,1] & \cdots&r_k[1,N] \\
    \vdots & \ddots&\vdots \\
    r_k[M,1] &\cdots& r_k[M,N] \\
  \end{array} } \right].
\end{align}


As mentioned previously, each task can be computed either by its own user's device or be offloaded to one of the servers.
In order to indicate such MEC strategy of user $k$ including channel access, define the $M\times N$ dimensional binary indicator matrix of user $k$ as
\begin{align}
\mathbf{S}_k=
  \left[ {\begin{array}{ccc}
    s_{k}[1,1] & \cdots&s_k[1,N] \\
    \vdots & \ddots&\vdots \\
    s_k[M,1] &\cdots& s_k[M,N] \\
  \end{array} } \right],
\end{align}
where $s_k[m,n]\in \{0,1\}$. Here, $s_k[m,n]=1$ means that user $k$ accesses server $m$ using channel $n$ in order to upload its input data of size $\alpha_k$. After receiving the input data from user $k$, server $m$ initiates its computation for task $J_k$, which will be stated in detail in Section \ref{subsec:computation_model}.

In this paper, we assume that each task is atomic so that it should be entirely assigned to a single server for computation offloading and cannot be partitioned and assigned to multiple servers. Hence, $\sum_{m\in[1:M],n\in[1:N]}s_k[m,n]\leq 1$ should be satisfied. 
Note that $\sum_{m\in[1:M],n\in[1:N]}s_k[m,n]=0$ means local computing. That is, user $k$ computes its task utilizing its own device.

\begin{figure}[]
	\centering
	\includegraphics[scale=0.68]{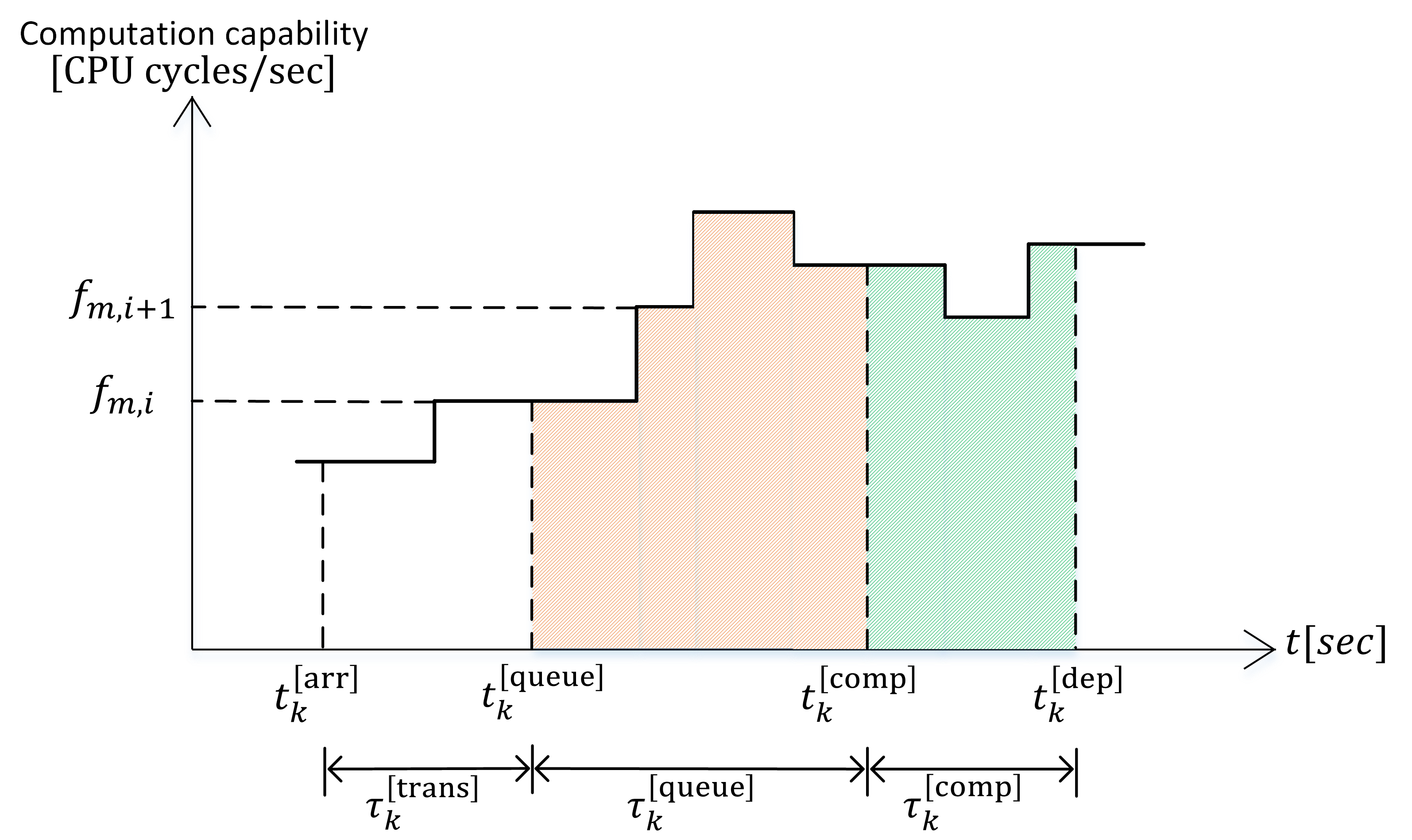}
	\centering
	\caption{Computational capability over time.}
	\label{fig:freq_variation}
\end{figure}

\subsection{Computation Model} \label{subsec:computation_model}
The primary aim of this paper is to minimize the delay for completing each task. Denote $t^{[\text{dep}]}_k$ as the time for completing $J_k$. 
We assume that the computation capability of user $k$ is given by $f^{[\text{user}]}_k$ [CPU cycles/sec].
Unlike fixed computation capability at the user side, we model the computation capability of server $m$ as a general time-varying function of time $t$, denoted by $f_m(t)$ as shown in Fig. \ref{fig:freq_variation}. To capture dynamics of each server's computation capability, we introduce a series of discrete update times of server $m$, denoted by $\{t^{[\text{up}]}_m[i]\}_{i\in\{1,2,\cdots\}}$. The computation capability of server $m$ at time $t^{[\text{up}]}_m[i]$ is renewed as $f_m(t^{[\text{up}]}_m[i])$ and then is remained as the same value until the next update time $t^{[\text{up}]}_m[i+1]$. For notational simplicity, denote $f_{m,i}=f_m(t^{[\text{up}]}_m[i])$.

\begin{remark}
Because each server typically deals with several different classes of tasks, the amount of computation resources that can be assigned to a single class of tasks might be changed over time.
Hence, in this paper, we assume a general time-varying computation capability for each server, which contains the conventional fixed computation capability model as a special case.
\hfill$\lozenge$
\end{remark}

\subsubsection{Local computing}
For local computing, i.e., $\sum_{m\in[1:M],n\in[1:N]}s_k[m,n]=0$, user $k$ will compute $J_k$ utilizing its own computing device.
Then the delay for completing $J_k$ is given by
\begin{align}\label{eq:local_delay}
\tau^{[\text{local}]}_k=\frac{\beta_k}{f^{[\text{user}]}_k}.
\end{align}
It is worthwhile to mention that the transmission delay due to uploading its input data becomes zero since the input data is already available for the case of local computing. 

\subsubsection{Edge computing}
Suppose that $\sum_{m\in[1:M],n\in[1:N]}s_k[m,n]=1$.
For this case, user $k$ firstly uploads its input data to one of the servers using an available channel.
Then the received input data will be stored at the queue memory of the server and then be served based on a FIFO manner.
Therefore, the queuing delay can be incurred before starting to compute $J_k$ due to the previously assigned tasks to the server.
Let $\tau^{[\text{queue}]}_k$  and $\tau^{[\text{comp}]}_k$ denote the queuing delay and the computation delay for $J_k$, respectively. 
For notational convenience, denote $t_k^{[\text{queue}]}$ as the time to complete uploading the input data to the queue memory and $t_k^{[\text{comp}]}$ as the time for starting the computation for $J_k$. 
Then, from the definitions, we have 
\begin{align} \label{eq:delay_definition}
t_k^{[\text{queue}]}&=t_k^{[\text{arr}]}+\tau^{[\text{trans}]}_k,\nonumber\\
t_k^{[\text{comp}]}&=t_k^{[\text{arr}]}+\tau^{[\text{trans}]}_k+\tau^{[\text{queue}]}_k,\nonumber\\
t_k^{[\text{dep}]}&=t_k^{[\text{arr}]}+\tau^{[\text{trans}]}_k+\tau^{[\text{queue}]}_k+\tau^{[\text{comp}]}_k.
\end{align} 
That is, the overall delay for completing $J_k$ is given by 
\begin{align} \label{eq:_delay_edge}
\tau_k^{[\text{MEC}]}=\tau^{[\text{trans}]}_k+\tau^{[\text{queue}]}_k+\tau^{[\text{comp}]}_k.
\end{align}
From \eqref{eq:uploading_rate}, the transmission delay for uploading the input data of size $\alpha_k$ is given by
\begin{align} \label{eq:tau_trans}
\tau^{[\text{trans}]}_k=\frac{\alpha_k}{\sum_{m,n}s_k[m,n]r_k[m,n]},
\end{align}
which is determined by $(\alpha_k, \mathbf{R}_k,  \mathbf{S}_k)$. 
On the other hand, the strategies of other tasks will obviously affect to $\tau^{[\text{queue}]}_k$ and, as a consequence, it is determined by $(\{t^{[\text{queue}]}_l\}_l, \{\beta_l\}_l,\{f_{m,i}\}_{m,i}, \{\mathbf{S}_l\}_l)$.
Finally, $\tau^{[\text{comp}]}_k$ is determined by $(t^{[\text{comp}]}_k, \beta_k, \{f_{m,i}\}_{m,i},\mathbf{S}_k)$.
Intuitively, $\tau^{[\text{queue}]}_k$ is determined by the time duration such that the integrated area in Fig.  \ref{fig:freq_variation} becomes the same as the remaining amount of computation at time $t_k^{[\text{queue}]}$. Similarly, $\tau^{[\text{comp}]}_k$ is determined by the time duration such that the integrated area in Fig.  \ref{fig:freq_variation} becomes the same as $\beta_k$.
The detailed derivation of how to calculate $\tau^{[\text{queue}]}_k$ and $\tau^{[\text{comp}]}_k$ is given in Appendix.

\begin{remark}
In this paper, we assume widely adapted network and computation models used in the literature, for instance, see \cite{DaixingxiaTCOD2DMEC,Ming_DRLoffloadqueue,xuchenEMTORAMEC,LiuLeiMAMHTOADV} and the references therein. Typical applications for such MEC systems include edge computing for vehicles or UAVs via 5G or IoV networks~\cite{surveymec,CaoBinLSMODOES,Livehicular,EiNwayEERAMUAV,zhaouav,DuanweiET5GIoVN}. Edge computing for industrial IoT applications~\cite{DaixingxiaTCOD2DMEC,CaoDRLJointaccessoffload,QiuTieECIITAA,LiuYagTEIMEC5GIoT} or augmented reality~\cite{chen_energyoffload}  has been also actively studied as key MEC applications.    \hfill$\lozenge$
\end{remark}

\subsection{Online Delay Minimization} \label{subsec:delay_min}
In this paper, we focus on an online optimization framework that is able to establish an MEC strategy of a newly arrived task in real time to minimize its own delay when the strategies of the previously arrived tasks are given.
To formally state such online delay minimization problem, let us define the $N$ dimensional binary indicator vector, which represents the status of $N$ channels of server $m$ at time $t$ as
\begin{align}
\mathbf{v}_m(t)=[v_{m,1}(t),v_{m,2}(t),\cdots, v_{m,N}(t)],
\end{align}
where $v_{m,n}(t)$ becomes one if channel $n$ of server $m$ is being used at time $t$ from one of the tasks appeared before time $t$ or becomes zero otherwise. 
That is, denote $\mathcal{I}(t)=\{l|t^{[\text{arr}]}_l\leq t\leq t^{[\text{queue}]}_l\}$ as the set of task indices that their input data are being uploaded at time $t$. Then
\begin{align}
v_{m,n}(t) = \begin{cases}
    1 & \text{if there is $l\in\mathcal{I}(t)$ such that} \\
      & s_l[m,n]=1, \\
    0 & \text{otherwise}.
\end{cases}
\end{align}

When the set of previous MEC strategies for $\{J_l\}_{l\in\{1:k-1\}}$ is given, i.e., $\{\mathbf{S}_l\}_{l\in[1:k-1]}$, the delay minimization for the MEC strategy of $J_k$ is represented by
\begin{align} \label{eq:tau}
&\min_{\mathbf{S}_k\in\{0,1\}^{M\times N}} \Big\{\mathbf{1}_{(\sum_{m,n}s_k[m,n]=0)}\tau^{[\text{local}]}_k  \nonumber\\
&{~~~~~~~~~~~~~~~~~~}+\mathbf{1}_{(\sum_{m,n}s_k[m,n]=1)}\tau_k^{[\text{MEC}]}\Big\}\\
&{~~~}\mbox{subject to} {~~}\sum_{m\in[1:M], n\in[1:N]}s_k[m,n]\leq 1. \label{eq:condition2}
\end{align}

In this paper, we focus on the delay minimization by properly choosing an MEC strategy of each arrived task, given by the optimization problem in \eqref{eq:tau} to \eqref{eq:condition2}. Let $\tau_k$ denote the delay of $J_k$ attained by solving the above optimization. Then the average delay is given by $\limsup_{K\to \infty} \frac{1}{K}\sum_{k=1}^K \tau_k$.

\section{Hybrid Online--Offline Task Offloading} \label{sec:rl}
In this section, we propose a hybrid online--offline learning based task offloading policy for minimizing the average delay experienced by tasks in the network. Because of network dynamics and time-varying computation capability of edge servers, it is crucially important to estimate the overall task delay when a task is offloaded to a certain edge server.
For this purpose, we firstly design a delay estimator which is implemented by a deep neural network (DNN) and is trained in an offline manner to accurately predict the overall task delay.
Then such a trained DNN is utilized as a core component for constructing a task offloading policy, which sequentially determines the computation strategy of each arrived task in an online manner in real time via a DRL agent.

\subsection{Delay Prediction for Edge Computing} \label{subsec:delay+prediction}
In this subsection, we firstly design a DNN-based delay estimator, which predicts the overall delay of each task when it is offloaded to a certain server by utilizing several useful information such as historical computation capabilities of the server, the amount of required CPU cycles for the assigned tasks, information about the current task, and current channel qualities for accurate estimation. We then explain how to train such DNN-based delay estimator via supervised learning.      
In particular, the proposed delay prediction method for edge computing consists of three components: data collection, offline training, and inference. A fully connected feed-forward DNN is used for this purpose. For training this neural network, training data is firstly collected during the online computation task offloading executed by the DRL agent. The detailed procedure of how to collect training data for the supervised learning will be given in Section \ref{subsec:online_policy}.

Suppose that $J_k$ is offloaded to server $m\in[1:M]$ by the DRL agent at time $t_k^{[\text{arr}]}$.
Then a training sample represented by the input $X$ and output $Y$ can be obtained.
Specifically, let  $\mathbf{f}_m(t_{k}^{[\text{arr}]}) \in \mathbb{R}^{U \times 1}$ be the vector containing the current and historical computation capabilities of server $m$ at time $t_{k}^{[\text{arr}]}$, where $U$ denotes the window size used to leverage historical information on computation capability for delay estimation.  Define $i^*$ such that $t_{m}^{[\text{up}]}[i^*]<t_{k}^{[\text{arr}]}<t_{m}^{[\text{up}]}[i^*+1]$. We have 
\begin{align}\label{eq:hist_freq}
\mathbf{f}_m(t_{k}^{[\text{arr}]})=\big[ f_{m,i^*}, f_{m,i^*-1}, ...,f_{m,i^*-U+1} \big].
\end{align}
Let $q_m(t)$ be the amount of required CPU cycles for the assigned tasks of server $m$ at time $t$.
We construct 
\begin{align} \label{eq:input_X}
X&=\left(\mathbf{f}_m(t_{k}^{[\text{arr}]}), q_{m}(t_{k}^{[\text{arr}]}), \alpha_{k}, \beta_{k}, r_k[m,n_m^*]\right),
\end{align}
where
\begin{align} \label{eq:optimal_n}
n_{m}^{*}=\underset{n \in [1:N],\text{ such that } v_{m,n}(t^{[\text{arr}]}_k)=0}{\arg \max} r_k[m,n],
\end{align}
which is the channel index with the highest rate among the available channels of server $m$ at time $t_k^{[\text{arr}]}$.

We construct $Y=\tau_k^{[\text{MEC}]}$. 
Note that $\tau_k^{[\text{MEC}]}$ is the actual delay in \eqref{eq:_delay_edge} that can be obtained after completing the computation of $J_k$.

Hence, the input layer size of the DNN is the same as the size of $X$ in \eqref{eq:input_X} and the output layer of the DNN is given as a scalar value, denoted by $g(X)$, which represents the estimated delay of $J_k$.
We train the DNN based on the set of such training samples in an offline manner.
Because the considered delay prediction is a regression problem, we apply supervised learning and adopt the mean squared error (MSE) as its loss function, given by
\begin{align}\label{eq:loss_function}
    \frac{1}{E}\sum_{i=1}^{E}(Y_{i}-g(X_{i}))^2,
\end{align}
where $E$ denotes the number of training samples used for defining the loss function, $(X_i,Y_i)$ denotes the $i$th training sample, and $g(X_i)$ denotes the produced output value of the DNN when $X_i$ is given.

\begin{figure}[]
    \centering
	\includegraphics[scale=0.68]{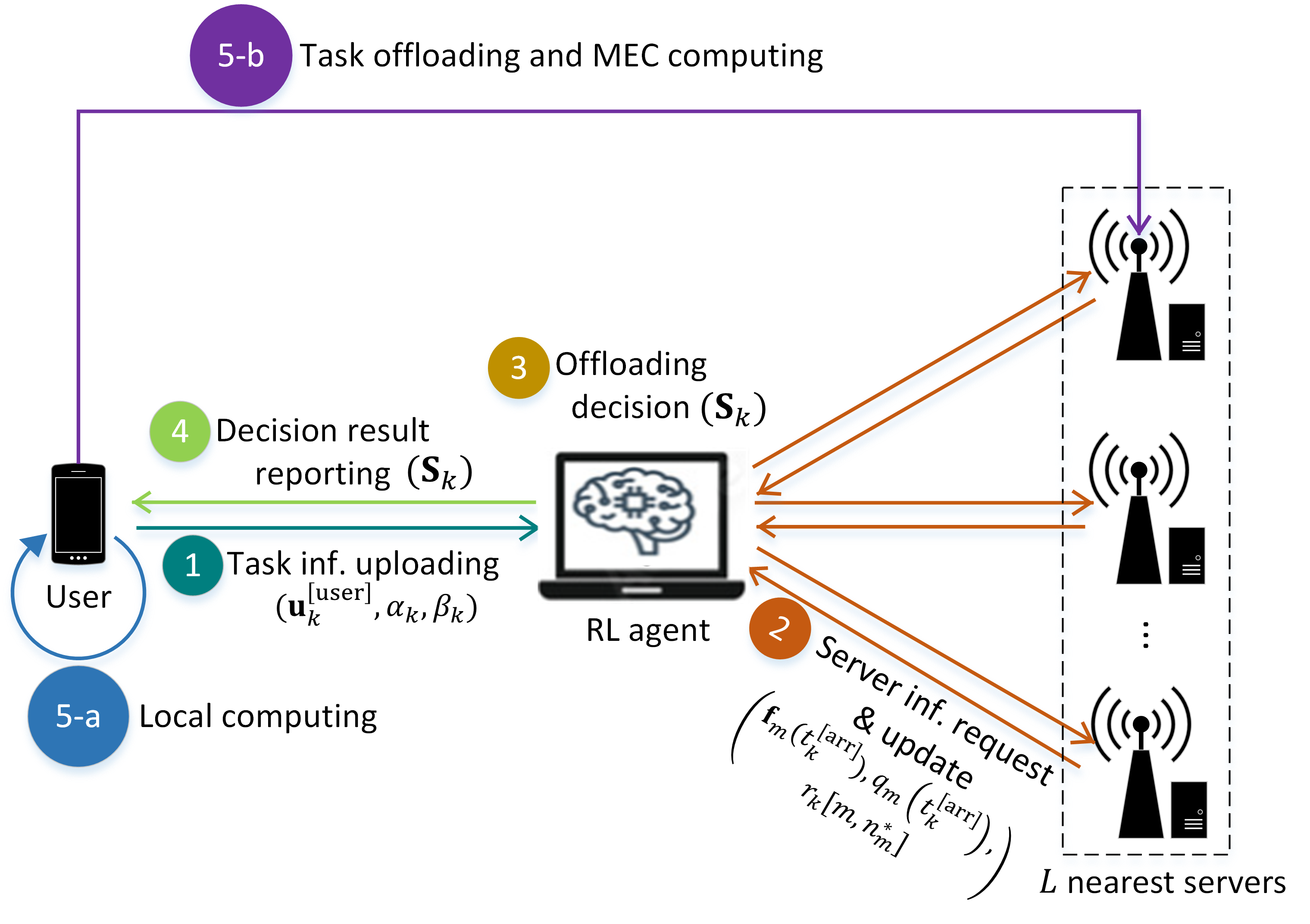}
	\centering
	\caption{Offloading decision procedure at the DRL agent.}
	\label{fig:offld_decis_proc}
\end{figure}

\subsection{Online Task Offloading Policy} \label{subsec:online_policy}
As mentioned in Section \ref{Network_Model}, users arrive  uniformly at random in the network area $[0,a]^2$ with arrival rate $\lambda$ [the average number of arrivals/sec] and generate tasks that need to be computed. 
In order to efficiently handle such randomly arrived tasks, we introduce a DRL agent that gathers necessary information from $L$ nearest server for determining the task offloading policy sequentially in real time when each user arrives.
Specifically, we propose a generalized edge computing framework that can efficiently handle randomly arrived computation tasks in real time under practical signaling and local information
reporting from edge servers as shown in Fig. \ref{fig:offld_decis_proc}.
For each task $J_{k}$, the task offloading policy is composed of series of steps as follows:
\begin{itemize}
    \item Step 1) User $k$ sends its location and task information, i.e., $\mathbf{u}_k^{[\text{user}]}$, $\alpha_k$, and $\beta_k$, to the DRL agent.
    \item Step 2) Let $\mathcal{O}_k$ be the set of $L$ nearest servers from $\mathbf{u}_k^{[\text{user}]}$. The DRL agent requests $\mathbf{f}_m(t_{k}^{[\text{arr}]})$, $q_m(t_k^{[\text{arr}]})$, and $r_k[m,n_m^*]$ to server $m\in\mathcal{O}_k$ and each server sends the requested information to the DRL agent, where the definitions of $\mathbf{f}_m(t_{k}^{[\text{arr}]})$, $q_m(t_k^{[\text{arr}]})$, and $r_k[m,n_m^*]$ are given in Section \ref{subsec:delay+prediction}.  
    \item Step 3) Based on the information delivered from user $k$ in Step 1 and from each server $m\in\mathcal{O}_k$ in Step 2, the DRL agent sets the MEC strategy of user $k$ by constructing $\mathbf{S}_k\in\{0,1\}^{M\times N}$ that satisfies the constraint in \eqref{eq:condition2}. The detailed procedure will be given later.
    \item Step 4) The DRL agent reports the resulting strategy $\mathbf{S}_k$ determined in Step 3 to user $k$.
    \item Step 5-a) If $\sum_{m\in[1:M],n\in[1:N]}s_k[m,n]=0$, user $k$ computes $J_k$ by utilizing its own computing device. Hence, the delay of $J_k$ is given by $\tau^{[\text{local}]}_k$ as in \eqref{eq:local_delay}.
    \item Step 5-b) If $s_k[m,n_m^*]=1$ for some $m\in\mathcal{O}_k$, user $k$ firstly uploads its input data to server $m\in\mathcal{O}_k$ using channel $n_m^*$ of server $m$. Then the received input data is stored at the queue memory and then be served based on a FIFO manner. Hence, the delay of $J_k$ is given by $\tau^{[\text{MEC}]}_k$ as in \eqref{eq:_delay_edge}.
\end{itemize}

Note that after Steps 1 and 2, $\mathbf{u}_k^{[\text{user}]}$, $\alpha_k$, $\beta_k$, $\mathbf{f}_m(t_{k}^{[\text{arr}]})$, $q_m(t_k^{[\text{arr}]})$, and $r_k[m,n_m^*]$ are available at the DRL agent, which are used for the offloading decision.
One possible implementation of the above framework is based on 5G systems, in which edge servers are equipped at BSs and the DRL agent is located at a BS controller (BSC) or mobile switching center (MSC). Hence, the required information between users and the DRL agent (Steps 1 and 4 in Fig. \ref{fig:offld_decis_proc}) can be delivered via wireless links between users and their serving BSs and wired backhaul links between BSs and the BSC. For information exchange between the DRL agent and edge servers (Step 2 in Fig. \ref{fig:offld_decis_proc}), wired backhaul links can be used.

\begin{algorithm}[!t] 
\caption{State generation of the DRL agent.}\label{alg:state_gener}
\begin{algorithmic}[1]
\State{{\bf Input}: $\mathbf{u}_k^{[\text{user}]}$, $\alpha_k$, $\beta_k$, $\{\mathbf{f}_m(t_{k}^{[\text{arr}]})\}_{m\in \mathcal{O}_k}$, $\{q_m(t_k^{[\text{arr}]})\}_{m\in \mathcal{O}_k}$, $\{r_k[m,n_m^*]\}_{m\in \mathcal{O}_k}$, and $\mathbf{a}_{k}=[a_{k-1},a_{k-2}, \cdots,a_{k-A}]$.}
\State{{\bf Initialization}: Set $\hat\Gamma_{k}^{[\text{MEC}]}=\mathbf{0}_{M}$, and $  \Omega_{k}=\mathbf{0}_{M}$.}
\For{all $m \in \mathcal{O}_k$}
\SplitState{Set $\big( \mathbf{f}_m(t_{k}^{[\text{arr}]}), q_{m}(t_{k}^{[\text{arr}]}), \alpha_{k}, \beta_{k}, r_k[m,n_m^*]\big)$ as the input of the delay estimator and obtain the estimated delay  $\hat\tau_{k,m}^{[\text{MEC}]}$, which corresponds to the output value of the estimator.}
\State{Update $[\hat\Gamma_{k}^{[\text{MEC}]}]_m \leftarrow \hat\tau_{k,m}^{[\text{MEC}]}$.}
\EndFor
\For{$j \in [1:L]$}
\State{Set $m'=\underset{m \in \mathcal{O}_k}{\arg \min}{[\hat\Gamma_{k}^{[\text{MEC}]}]_{m}}.$}
\State{Update $[\Omega_{k}]_{m'} \leftarrow j.$}
\State{Update $\mathcal{O}_k \leftarrow \mathcal{O}_{k}\setminus \{m'\}$.}
\EndFor
\State{{\bf Output}: $\mathcal{X}_k = (\Omega_k, \mathbf{a}_{k}$).}
 \end{algorithmic}
\end{algorithm}

We now explain the proposed online task offloading strategy stated in Step 3 in details.
Denote the state, action, and reward for task $J_k$ by $\mathcal{X}_k$, $a_k$, and $\Lambda_k$, which will be utilized at the DRL agent. 
Algorithm \ref{alg:state_gener} describes the construction of $\mathcal{X}_k$, where $\mathbf{0}_M$ denotes the all-zero vector of size $M$ and $[\mathbf{A}]_m$ denotes the $m$th element of a vector $\mathbf{A}$.
Note that $(\mathbf{u}_k^{[\text{user}]}, \alpha_k,\beta_k)$ is available after Step 1, $\{\mathbf{f}_m(t_{k}^{[\text{arr}]}), q_m(t_k^{[\text{arr}]}), r_k[m,n_m^*]\}_{m\in \mathcal{O}_k}$ is available after Step 2, and $\mathbf{a}_{k}=[a_{k-1},a_{k-2}, \cdots,a_{k-A}]$ is the historical information of the $A$ recently selected actions, where $A$ denotes a predetermined positive integer. Hence, all input values in Line 1 is available at the DRL agent.
In Line 4, we apply the delay estimator given in Section \ref{subsec:delay+prediction} with the set of inputs as in \eqref{eq:input_X} to estimate the delay of edge computing for every server $m \in \mathcal{O}_k$. Next, we store in $\Omega_{k}$ the ranking of the $L$ servers having the shortest delays as part of the state in Line 7 to 11. Finally, we construct $\mathcal{X}_{k}$ as a combination of $\Omega_k$ and the set of historical actions taken so far as in Line 12.

As mentioned previously, the role of the DRL agent is to determine local or edge computing of $J_k$ among the $L$ nearest servers in $\mathcal{O}_k$ for $k=1,2,\cdots$. For this purpose, the DRL agent utilizes a deep Q-network (DQN) to estimate the Q-value of each action. 
Note that the action space of $J_k$ is given by $\mathcal{A}_{k}= \{ 0 \}\cup\mathcal{O}_{k}$.
Therefore, the set $\mathcal{O}_{k}$ depends on each task $J_k$ so that the execution and training based on a common DQN cannot be straightforwardly given. To handle such a time-varying action space, we firstly define the sets $\hat\Gamma_{k}^{[\text{MEC}]}$ and $\Omega_{k}$ as the $M$-dimensional vectors in Algorithm \ref{alg:state_gener} and, as a result, the input layer of the common DQN is given as $\mathcal{X}_k=(\Omega_k, \mathbf{a}_{k})$ and the $m$th element in $\Omega_k$ corresponds to server $m$ for $m\in[1:M]$. 
For the servers that are not included in $\mathcal{O}_{k}$, we set zeros in $\Omega_k$.
Similarly, we define the output layer of the common DQN as the $(M+1)$-dimensional vector such that the $(m+1)$th element corresponds to the estimated Q-value of server $m$ for $m\in[1:M]$. The first element denotes the estimated Q-value for local computing.

The DRL agent applies the constructed $\mathcal{X}_k$ from Algorithm \ref{alg:state_gener} to the DQN and obtain $M+1$ estimated Q-values. Then the action of the DRL agent for task $J_k$ is determined by the maximum of the estimated Q-values in $\mathcal{A}_k$.
Finally, from the selected $a_k\in\mathcal{A}_{k}$, the MEC strategy $\mathbf{S}_k$ is given by
$s_k[a_k,n_{a_k}^{*}]=1$ and $s_k[m,n]=0$ for all $m\in[1:M]$ and $n\in[1:N]$ except the case where $(m,n)=(a_k,n_{a_k}^{*})$ when $a_k \neq0$. On the other hand, $s_k[m,n]=0$ for all $m\in[1:M]$ and $n\in[1:N]$ when $a_k=0$.
The reward of  $J_k$ is defined as
\begin{equation}\label{eq:reward}
\Lambda_{k}=-\mathbf{1}_{(\sum_{m,n}s_k[m,n]=0)}\tau^{[\text{local}]}_k-\mathbf{1}_{(\sum_{m,n}s_k[m,n]=1)}\tau_k^{[\text{MEC}]},
\end{equation}
which corresponds to the minus value of the delay of $J_k$.

\begin{figure}[t!]
	\centering
	\includegraphics[scale=0.66]{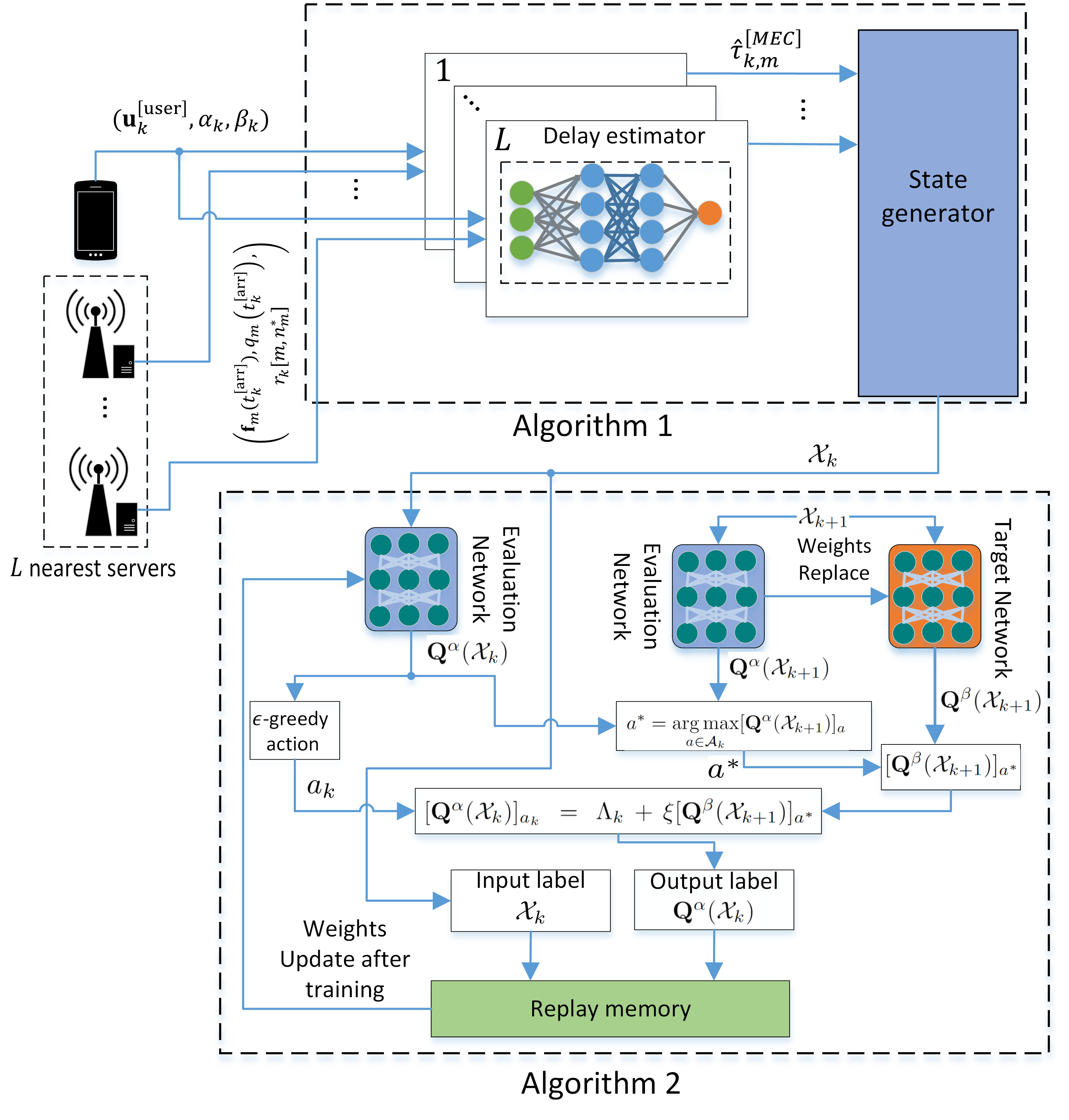}
	\centering
        \caption{Proposed DRL agent algorithm and training methodology.}
	\label{fig:RL_agent}
\end{figure}

\subsection{DRL Agent Operation and Training Methodology} \label{subsec:q_learning}
For the considered task offloading framework, a DRL algorithm namely double DQN \cite{van2016deep}, which is an extension of the conventional DQN algorithm, has been adopted. It is a model-free algorithm, i.e., can provide solutions to complex DRL problems without requiring a transition probability distribution. With the state, action, and reward functions defined in section \ref{subsec:online_policy}, we now describe the double DQN method.

Note that double DQN is useful for a reinforcement learning algorithm that utilizes two separate neural networks to improve the accuracy and stability of Q-value estimates. Therefore, the DRL agent consists of two DQNs, i.e., $\mbox{DQN}^{\alpha}$ and $\mbox{DQN}^{\beta}$, called the evaluation network and the target network, respectively. In this paper, the identical feed-forward fully connected neural network is used for both $\mbox{DQN}^{\alpha}$ and $\mbox{DQN}^{\beta}$, each of which generates Q-values. For each task $J_k$, the first network $\mbox{DQN}^{\alpha}$ is used for choosing and executing an action based on $\epsilon$-greedy policy. The second network $\mbox{DQN}^{\beta}$ is used to estimate the corresponding Q-values associated with the selected action. The target network is periodically updated with the weights of the evaluation network to ensure that the Q-values are accurate and stable over time. This allows double DQN to effectively handle non-stationary environments and overcome the problem of overestimating Q-values that is commonly seen in traditional Q-learning algorithms.

More specifically, the evaluation Q-network $\mbox{DQN}^{\alpha}$ in the double DQN algorithm is responsible for selecting the action to take in the environment and is updated during the learning process. The update of the $\mbox{DQN}^{\alpha}$ is performed using the standard Q-learning update rule, which is a variant of the Bellman equation. The Q-network is updated using the difference between the current estimate of the Q-value and the estimate of the Q-value obtained using the Bellman equation\cite{49}. The parameters in the evaluation Q-network are updated with the goal of minimizing this difference. The update equation for the evaluation Q-network is given by
\begin{multline}\label{Q_update}
Q\left( {\mathcal{X},a} \right) \leftarrow \left( {1 - \rho } \right) Q\left( {\mathcal{X},a} \right)\\
+\rho \left[ R\left( {\mathcal{X},a} \right) + \xi Q\left( \mathcal{X}', \mathop {\arg\max}\limits_{a' \in\mathcal{A}} Q(\mathcal{X}',a') \right) \right],
\end{multline}
where $Q(\mathcal{X},a)$ is the current estimate of the Q-value for state $\mathcal{X}$ and action $a$. Here $\rho>0$ is the learning rate, and $\xi\in [0,1]$ is the discount factor that determines the impact of future rewards, $\mathcal{X}'$ is the next state determined by $\mathcal{X}$ and $a$, and $\mathop {\arg\max}\limits_{a' \in\mathcal{A}} Q(\mathcal{X}',a')$ is the action with the highest Q-value in the next state as predicted by the evaluation network. This update process is iteratively applied over many episodes of interaction with the environment to improve the accuracy of the evaluation network.
Algorithm \ref{Algorithm2} describes the detailed DRL agent operation including the training methodology. 
Denote the set of weights in $\mbox{DQN}^{\alpha}$ by $\theta^{\alpha}$ and the set of weights in  $\mbox{DQN}^{\beta}$ by $\theta^{\beta}$, which are initialized as predetermined values.
For each task $J_k$, the state $\mathcal{X}_k$ (constructed from Algorithm \ref{alg:state_gener}) is fed into the evaluation network $\mbox{DQN}^\alpha$ and $M+1$ Q-values are generated, i.e.,
\begin{multline}
    \mathbf{Q}^{\alpha}(\mathcal{X}_k;\mathbf{\theta}^{\alpha})=[Q^{\alpha}(\mathcal{X}_k, a_k=0;\mathbf{\theta}^{\alpha}),Q^{\alpha}(\mathcal{X}_k, a_k=1;\mathbf{\theta}^{\alpha}),\\ \cdots,Q^{\alpha}(\mathcal{X}_k, a_k=M;\mathbf{\theta}^{\alpha})].
\end{multline}
For notational simplicity, we assume $a_k\in[0,M]$ in the update equation of Q-values. As stated in Section \ref{subsec:online_policy}, $a_k$ is actually chosen in $\mathcal{A}_{k}= \{ 0 \}\cup\mathcal{O}_{k}$ for online task offloading.
In order to make an optimal action decision, the DRL agent adopts the $\epsilon$-greedy strategy as seen in Line 5 of Algorithm \ref{Algorithm2}.
After the DRL agent takes the action $a_k$, it will receive the corresponding reward $\Lambda_k$ as in \eqref{eq:reward} and the next state $\mathcal{X}_{k+1}$ will be given, which can be obtained by performing Algorithm \ref{alg:state_gener}. The next state $\mathcal{X}_{k+1}$ is then fed into both the evaluation and target networks, and the corresponding Q-values generated from them are given as
\begin{multline}
        \mathbf{Q}^{\alpha}(\mathcal{X}_{k+1};\mathbf{\theta}^{\alpha})=[Q^{\alpha}(\mathcal{X}_{k+1}, a_{k+1}=0;\mathbf{\theta}^{\alpha}),\cdots,\\ Q^{\alpha}(\mathcal{X}_{k+1}, a_{k+1}=M;\mathbf{\theta}^{\alpha})]
        \end{multline}
and
        \begin{multline}
            \mathbf{Q}^{\beta}(\mathcal{X}_{k+1};\mathbf{\theta}^{\beta})=[Q^{\beta}(\mathcal{X}_{k+1}, a_{k+1}=0;\mathbf{\theta}^{\beta}),\cdots,\\ Q^{\beta}(\mathcal{X}_{k+1}, a_{k+1}=M;\mathbf{\theta}^{\beta})],
\end{multline}
respectively. In the next step, $\mathbf{Q}^{\alpha}(\mathcal{X}_k)$ is updated as in Line  8 of Algorithm \ref{Algorithm2} and each state $\mathcal{X}_k$ and the corresponding updated Q-values are stored in the experience replay memory. The evaluation network $\mbox{DQN}^{\alpha}$ is then trained using the data samples in the replay memory. The MSE metric is adopted as a loss function given by $\frac{1}{B}\sum_{i=1}^{B}\left\|\mathbf{Q}^{\alpha}(\mathcal{X}^{(i)})-\hat{\mathbf{Q}}^{\alpha}(\mathcal{X}^{(i)})\right\|^2$,
where $B$ is the size of mini-batch used for training $\mbox{DQN}^{\alpha}$ and $(\mathcal{X}^{(i)},\mathbf{Q}^{\alpha}(\mathcal{X}^{(i)}))$ are the $i$th data sample.
Also, $\hat{\mathbf{Q}}^{\alpha}(\mathcal{X}_k)$ is the estimated Q-values generated by $\mbox{DQN}^{\alpha}$ utilizing $\mathcal{X}^{(i)}$ as its input. Finally, the set of weights $\mathbf{\theta}^{\beta}$ is replaced with the set of updated weights $\mathbf{\theta}^{\alpha}$ as seen in Line 11 of Algorithm \ref{Algorithm2}.
For each arrived task $J_k$, the above process is repeated to make an offloading decision and training at the same time in an online manner.
Fig. \ref{fig:RL_agent} illustrates the proposed Q-learning and training methodology.

\begin{algorithm}[!ht] 
\caption{DRL agent operation.}\label{Algorithm2}
\begin{algorithmic}[1]
\State{{\bf Initialization}: Set the weights in $\mbox{DQN}^{\alpha}$ and $\mbox{DQN}^{\beta}$ as predetermined initial values.}
\For{each task $ J_{k}$ at $t_{k}^{[\text{arr}]}$}
\SplitState{Construct the state $\mathcal{X}_k$ of the DRL agent from the steps in Algorithm \ref{alg:state_gener}.}
\SplitState{The state $\mathcal{X}_{k}$ is fed into $\mbox{DQN}^{\alpha}$ and the corresponding Q-values $\mathbf{Q}^{\alpha}(\mathcal{X}_k)$ are obtained.}
\SplitState{The DRL agent determines $a_{k}$ using the $\epsilon$-greedy strategy. Specifically, $a_{k}={\arg\max }_{a\in\mathcal{A}_{k}}[\mathbf{Q}^{\alpha}(\mathcal{X}_{k})]_{a}$ with probability $1-\epsilon$ or $a_{k}$ is chosen uniformly at random in $\mathcal{A}_{k}$ with probability $\epsilon$.}
\SplitState{After taking the action $a_{k}$ (computation of $J_k$), $\Lambda_{k}$ is given as in \eqref{eq:reward}.}
\SplitState{The next state $\mathcal{X}_{k+1}$ is fed into $\mbox{DQN}^{\alpha}$ and $\mbox{DQN}^{\beta}$, and the corresponding Q-values $\mathbf{Q}^{\alpha}(\mathcal{X}_{k+1})$ and $\mathbf{Q}^{\beta}(\mathcal{X}_{k+1})$ are obtained.}
\SplitState{Update $\mathbf{Q}^{\alpha}(\mathcal{X}_k)$ in Line 4 by utilizing $\mathbf{Q}^{\alpha}(\mathcal{X}_{k+1})$ and $\mathbf{Q}^{\beta}(\mathcal{X}_{k+1})$ in Line 7 as
$[\mathbf{Q}^{\alpha}(\mathcal{X}_{k})]_{a_{k}}={\Lambda}_{k}+\xi [\mathbf{Q}^{\beta}(\mathcal{X}_{k+1})]_{a^{*}}, \mbox{where } a^{*}={{\arg\max }_{a \in\mathcal{A}_{k}}[\mathbf{Q}^{\alpha}(\mathcal{X}_{k+1})]_{a}}.$}
\SplitState{Store $\mathcal{X}_{k}$ and the corresponding Q-values $\mathbf{Q}^{\alpha}(\mathcal{X}_{k})$ in the experience replay memory.}
\SplitState{By using data samples in the experience replay memory, train $\mbox{DQN}^{\alpha}$.}
\SplitState{Update the weights in $\mbox{DQN}^{\beta}$ with the weights in $\mbox{DQN}^{\alpha}$.}
\EndFor
 \end{algorithmic}
\end{algorithm}

\section{General Cost Functions and Partial Offloading} \label{sec:extension}
In this section, we discuss general cost functions and how our proposed online--offline task offloading method can be applied to such general cost functions. Then we also discuss how to extend our method to partial offloading model.

\subsection{General Cost Functions} \label{subsec:general_cost}
Previously, we focused on minimizing the overall delay of MEC systems. In the literature, several works have also considered energy consumption during edge computing~\cite{chen_energyoffload,
zmingx_EATORA}. With the explosive growth of intelligent internet of things (IoT) devices, most computation-intensive devices not only require ultra-low latency services but also energy-efficient computation services because of limited battery power. Therefore, minimizing costs related to both delay and energy are equally important. For this purpose, we introduce two types of cost functions in this section: the first is a delay-dependent cost function and the second is an energy-dependent cost function.

Denote $C^{[\text{d}]}_k$ and $C^{[\text{e}]}_k$ as the delay-dependent cost function and the energy-dependent cost function of task $J_k$, respectively. Let us first define $C^{[\text{d}]}_k$ when the delay of $J_k$ is given by $\tau_k$. Recall that $\tau_k=\mathbf{1}_{(\sum_{m,n}s_k[m,n]=0)}\tau^{[\text{local}]}_k+\mathbf{1}_{(\sum_{m,n}s_k[m,n]=1)}\tau_k^{[\text{MEC}]}$ from \eqref{eq:tau}.
Let $C^{[\text{d}]}_k=C(\tau_k) \in [0, C_{\max}]$ be the cost associated with the delay $\tau_k$. 
Obviously, $C(\tau)$ should be a non-increasing function with respect to $\tau>0$. To prevent the case where the cost of a specific task becomes excessively large, we assume that the maximum cost of each task is limited by $C_{\max}$.
Depending on different service types, we mainly focus on the following classes of cost functions.
\begin{itemize}
\item Strict delay services: $C(\tau)=0$ for $\tau\leq \tau_{\text th}$ and $C(\tau)=C_{\max}$ otherwise, where $ \tau_{\text th}>0$ is the deadline.
\item Delay-sensitive services: $C(\tau)=\min (\exp(c_1 \tau)-1,C_{\max})$ or $C(\tau)=\min (\tau^{c_2},C_{\max})$, where $c_1>0$ and $c_2>1$.
\item Delay-tolerant services: $C(\tau)=\min (\tau^{c_3},C_{\max})$ or $C(\tau)=\min (\log(1+c_4\tau),C_{\max})$, where $c_3\leq 1$ and $c_4>0$.
\end{itemize}

Let us now define the energy-dependent cost function of $J_k$, denoted by $C^{[\text{e}]}_k$. 
According to the widely adopted model for energy consumption \cite{Energy_cal_chen_xu}, the energy consumed in computing $J_k$ with the required amount of computation $\beta_k$\ is obtained for the case of local computing, i.e., $\sum_{m,n}s_{k}[m,n]=0$, as 
\begin{equation}
    E_{k}^{ \left[\text{local} \right]}=\kappa \left(f_{k}^{ \left[\text{user} \right]} \right)^{2} \beta_k.
\end{equation}
For the case of edge computing with server $m$, i.e., $\sum_{n\in[1:N]}s_{k}[m,n]=1$, similar approach in Appendix can be used for the consumed energy calculation. 
In order to calculate the consumed energy, we define $j''$ such that $t_m^{[\text{up}]}[j'']\leq t_k^{[\text{comp}]}\leq t_m^{[\text{up}]}[j''+1]$.
Let us first consider the case where $f_{m,j''}(t_m^{[\text{up}]}[j''+1]-t_k^{[\text{comp}]})>\beta_k$. For this case the consumed energy is given by
\begin{equation}
    E_{k,m}^{[\text{MEC}]}=\kappa (f_{m,j''})^{2} \beta_{k},
\end{equation}
where $\kappa =10^{-27}$ denotes the energy efficiency parameter that is mainly dependent on the chip architecture~\cite{Energy_cal_chen_xu}.
Otherwise, find a non-negative integer value $z''$ such that equation \eqref{eq:comp_data_req_f} is satisfied. Then the energy for this case is
\begin{align}
    E_{k,m}^{[\text{MEC}]}=&~\kappa \left(f_{m,j^{''}} \right)^{3} (t_{m}^{[\text{up}]}[j^{''}+1]-t_{k}^{[\text{comp}]}) \nonumber\\&+ \sum_{z=1}^{z^{''}}\kappa \left(f_{m,j^{''}+z} \right)^{3}(t_{m}^{[\text{up}]}[j^{''}+z+1]-t_{m}^{[\text{up}]}[j^{''}+z])\nonumber\\&+\kappa \left(f_{m,j^{''}+z^{''}+1} \right)^{2}(\beta_k-\bar{\beta}_k),
\end{align}
where $\bar{\beta}_{k}$ is defined as in \eqref{eq:bar_beta}.
Hence, we have 
\begin{align} 
C^{[\text{e}]}_k=\mathbf{1}_{(\sum_{m,n}s_k[m,n]=0)}E^{[\text{local}]}_k+\mathbf{1}_{(\sum_{m,n}s_k[m,n]=1)}E_k^{[\text{MEC}]}.
\end{align} 
Finally, the overall cost of $J_k$ is represented by the weighted sum of $C^{[\text{d}]}_k$ and $C^{[\text{e}]}_k$ as
\begin{align} \label{eq:cost}
C_k=\omega^{[\text{d}]} C^{[\text{d}]}_k +\omega^{[\text{e}]} C^{[\text{e}]}_k,
\end{align}
where $\omega^{[\text{d}]}\geq 0$ and $\omega^{[\text{e}]}\geq 0$ are the weights representing the contribution or importance of $C^{[\text{d}]}_k$ and $C^{[\text{e}]}_k$ to the overall cost.

\begin{algorithm}[!t] 
\caption{State generation for the general cost function.}\label{alg:state_gener_energy_delay}
\begin{algorithmic}[1]
\State{{\bf Input}: $\mathbf{u}_k^{[\text{user}]}$, $\alpha_k$, $\beta_k$, $\{\mathbf{f}_m(t_{k}^{[\text{arr}]})\}_{m\in \mathcal{O}_k}$, $\{q_m(t_k^{[\text{arr}]})\}_{m\in \mathcal{O}_k}$, $\{r_k[m,n_m^*]\}_{m\in \mathcal{O}_k}$, and $\mathbf{a}_{k}=[a_{k-1},a_{k-2}, \cdots,a_{k-A}]$.}
\State{{\bf Initialization}: Set $\hat\Gamma_{k}^{[\text{MEC}]}=\mathbf{0}_{M}$, $\hat\Upsilon_{k}^{[\text{MEC}]}=\mathbf{0}_{M}$, and $  \Omega_{k}=\mathbf{0}_{M}$.}
\For{all $m \in \mathcal{O}_k$}
\SplitState{Set $\big( \mathbf{f}_m(t_{k}^{[\text{arr}]}), q_{m}(t_{k}^{[\text{arr}]}), \alpha_{k}, \beta_{k}, r_k[m,n_m^*]\big)$ as the input of the delay estimator and obtain the estimated delay  $\hat\tau_{k,m}^{[\text{MEC}]}$, which corresponds to the output value of the delay estimator.}
\State{Update $[\hat\Gamma_{k}^{[\text{MEC}]}]_m \leftarrow \hat\tau_{k,m}^{[\text{MEC}]}$.}
\SplitState{Set $\big( \mathbf{f}_m(t_{k}^{[\text{arr}]}), q_{m}(t_{k}^{[\text{arr}]}), \alpha_{k}, \beta_{k}, r_k[m,n_m^*]\big)$ as the input of the energy estimator and obtain the estimated energy  $\hat E_{k,m}^{[\text{MEC}]}$, which corresponds to the output value of the energy estimator.}
\State{Update $[\hat \Upsilon_{k}^{[\text{MEC}]}]_m \leftarrow \hat E_{k,m}^{[\text{MEC}]}$.}
\EndFor

\For{$j \in [1:L]$}
\SplitState{Set $m'=\underset{m \in \mathcal{O}_k}{\arg \min}{\left[w^{[\text{d}]}\hat\Gamma_{k}^{[\text{MEC}]}+ w^{[\text{e}]}\hat\Upsilon_{k}^{[\text{MEC}]}\right]_{m}}$, where $C(\hat\Gamma_{k}^{[\text{MEC}]})$ denotes the element-wise operation, i.e., $\left[C(\hat\Gamma_{k}^{[\text{MEC}]})\right]_m= C\left([\hat\Gamma_{k}^{[\text{MEC}]}]_m\right)$, and $C(\cdot)$ denotes the delay-dependent cost function defined in Section \ref{subsec:general_cost}.}
\State{Update $[\Omega_{k}]_{m'} \leftarrow j.$}
\State{Update $\mathcal{O}_k \leftarrow \mathcal{O}_{k}\setminus \{m'\}$.}
\EndFor

\State{{\bf Output}: $\mathcal{X}_k = (\Omega_k, \mathbf{a}_{k}$).}
 \end{algorithmic}
\end{algorithm}
\subsection{Online Task Offloading Policy and Training Methodology}
\begin{figure}[t!]
	\centering
	\includegraphics[scale=0.66]{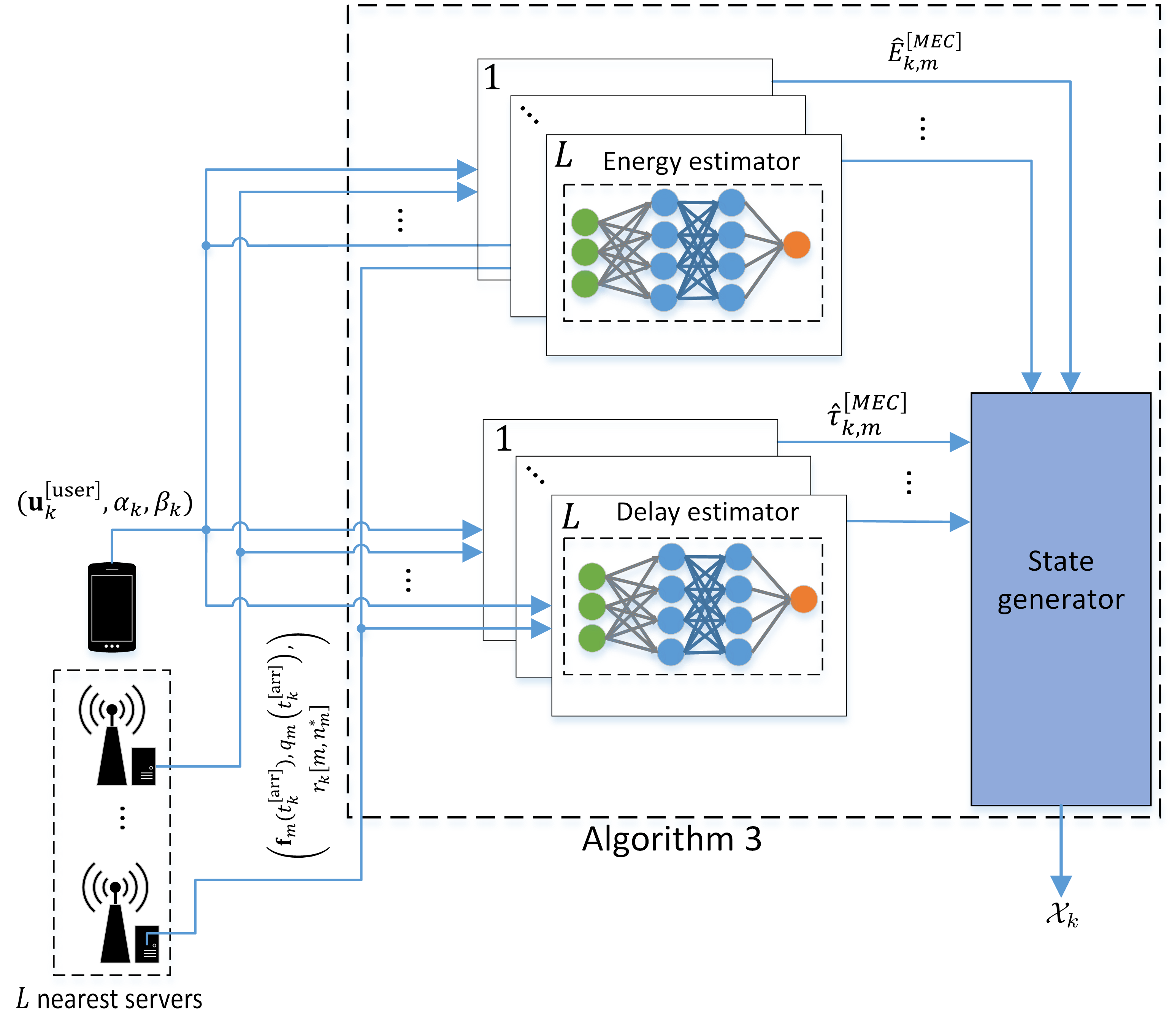}
	\caption{State generation based on delay and energy estimators.}
	\label{fig:delay and energy estimator}
\end{figure}

In order to enable the proposed scheme to minimize general cost functions defined in Section \ref{subsec:general_cost}, generalization and modification of the proposed scheme applicable for such general cost functions are required. A key difference is that in addition to the delay prediction for each task, prediction of the consumed energy in processing the task is also required.
Recall that the delay prediction method for edge computing was introduced in Section \ref{subsec:delay+prediction}, which is based on offline training via DNN. In a similar manner, another DNN-based estimator can be adopted for predicting the consumed energy.
Suppose that $J_k$ is offloaded to server $m\in [1:M]$ by the DRL agent at time $t^{ \left[\text{arr} \right]}_k$. Then a training sample that is represented by its input label $X$ and output label $Y$ can be obtained. The input label remains the same as in \eqref{eq:input_X}. We construct $Y= E^{ \left[\text{MEC} \right]}_{k}$. Note that $E^{ \left[\text{MEC} \right]}_{k}$ is the actual consumed energy after completing the computation of $J_k$. The DNN for energy prediction can be trained similarly as the DNN for delay prediction using the MSE as a loss function as given in \eqref{eq:loss_function}. The proposed online task offloading policy and training methodology for minimizing the weighted sum cost in \eqref{eq:cost} are similarly given as those in Sections \ref{subsec:online_policy} and \ref{subsec:q_learning}. The key difference is in the construction of state $\mathcal{X}_k$ used for the DRL agent.
Fig. \ref{fig:delay and energy estimator} illustrates how to construct $\mathcal{X}_k$ for the case of the weighted sum cost in \eqref{eq:cost}. 

Specifically, Algorithm \ref{alg:state_gener_energy_delay} describes the construction of $\mathcal{X}_k$.
As explained in Section \ref{subsec:online_policy}, all input values in Line 1 is available at the DRL agent.
In Line 4, we apply the delay estimator given in Section \ref{subsec:delay+prediction} to estimate the delay of edge computing for server $m \in \mathcal{O}_k$. 
In Line 6, we also apply the energy estimator to estimate the consumed energy of edge computing for server $m \in \mathcal{O}_k$. After constructing  $\Omega_{k}$ from $\hat\Gamma_{k}^{[\text{MEC}]}$ and $\hat\Upsilon_{k}^{[\text{MEC}]}$  in Line 9 to Line 13, we finally obtain $\mathcal{X}_k$ in Line 14. Apart from the state generation procedure, the remaining online task offloading policy and training methodology of the DRL agent will remain the same as discussed in Section \ref{subsec:q_learning}.

\begin{table}[h]
  \begin{center}
    \caption{Simulation parameters.}
    \label{tab:table1}
	\scalebox{0.95}{
    \begin{tabular}{l|c} 
     \hline 
	  \multicolumn{1}{c|}{Parameter} & Value\\
      \hline \hline
      \multicolumn{1}{c|}{Network area} & $[-5, 5] \times [-5, 5] {\text{ km}}^2$ \\
		\hline
      \multicolumn{1}{c|}{Number of MEC servers $(M)$} & 15\\
		\hline
		\multicolumn{1}{c|}{Bandwidth per server $(W)$} & 20 MHz\\
		\hline
		\multicolumn{1}{c|}{Number of wireless channels $(N)$} & $10$\\
       \hline
		\multicolumn{1}{c|}{Transmit power $(P_k)$} & $23$ dBm \\
		\hline
		\multicolumn{1}{c|}{Path loss exponent $(\gamma)$} & $3.8$\\
       \hline
		\multicolumn{1}{c|}{Task data size $(\alpha_k)$} & $[8,12] \times 10^{6}$ bits \\
		\hline
		\multicolumn{1}{c|}{Task computation size $(\beta_k)$} & $[7,8] \times 10^{9}$ CPU cycles \\
		\hline
		\multicolumn{1}{c|}{Thermal noise $(N_0)$} & $-174$ dBm/Hz \\
		\hline
		\multicolumn{1}{c|}{MEC computation capability} & $[5,12] \times 10^{9}$ CPU cycles/sec \\
		\hline
		\multicolumn{1}{c|}{Learning rate $(\rho)$}& $0.001$\\
		\hline
		\multicolumn{1}{c|}{Discount factor $(\xi)$} & $0.95$\\
		\hline
    \end{tabular}}
  \end{center}
\end{table}

\subsection{Partial Offloading}\label{subsec:partial_offload}
In this paper, we have focused on a binary task offloading policy in which each task is solely computed at its local device or completely offloaded to one of the nearest servers. However, the proposed online--offline offloading framework can be extended to partial offloading in order to further improve the delay performance by utilizing parallel computing at both user and edge server sides.
Let $v_k\in[0,1]$ be the offloading ratio for $J_k$.
Then the amount of input data and the required amount of computation assigned to one of servers are given by $v_k\alpha_k$ [bits] and $v_k\beta_k$ [CPU cycles], respectively. Hence, from \eqref{eq:local_delay}, the local computing delay for $J_k$ is given by
\begin{align}\label{eq:local_partial}
    \tau_{k}^{[\text{local}]}=\frac{(1-v_k)\beta_k}{f_{k}^{[\text{user}]}}.
\end{align}
Now consider the edge computing delay. 
From \eqref{eq:tau_trans}, the transmission delay for uploading the assigned input data is given by
\begin{align} \label{eq:tau_trans_partial}
\tau^{[\text{trans}]}_k=\frac{v_k\alpha_k}{\sum_{m,n}s_k[m,n]r_k[m,n]},
\end{align}
which is determined by $(v_k\alpha_k, \mathbf{R}_k,  \mathbf{S}_k)$. Also, as mentioned in Section \ref{subsec:computation_model}, $\tau^{[\text{queue}]}_k$ and $\tau^{[\text{comp}]}_k$ will be determined by $(\{t^{[\text{queue}]}_l\}_l, \{v_k\beta_l\}_l,\{f_{m,i}\}_{m,i}, \{\mathbf{S}_l\}_l)$ and $(t^{[\text{comp}]}_k, v_k\beta_k, \{f_{m,i}\}_{m,i},\mathbf{S}_k)$ respectively, and the same analysis in Appendix can be applied to calculate $\tau^{[\text{queue}]}_k$ and $\tau^{[\text{comp}]}_k$, which result in $\tau_k^{[\text{MEC}]}$.
Finally, suppose that $\mathcal{V}$ is the set of discrete offloading ratios.
Then, from \eqref{eq:tau} and \eqref{eq:condition2}, the delay minimization for the partial MEC strategy of $J_k$ is reformulated as 

\begin{align} \label{eq:tau_partial}
&\min_{\mathbf{S}_k\in\{0,1\}^{M\times N},v_k\in\mathcal{V}} \, \max \left\{ \tau^{[\text{local}]}_k, \tau_k^{[\text{MEC}]} \right\}\\
&{~~~~~}\mbox{subject to}{~~~~~~}\sum_{m\in[1:M], n\in[1:N]}s_k[m,n]= 1. \end{align}

In order to enable the proposed scheme for partial offloading, modification of the proposed scheme is required. Recall that the delay prediction method for edge computing is introduced in Section \ref{subsec:delay+prediction}, which is based on offline training via DNN. For partial offloading, the input label $X$ in \eqref{eq:input_X} is redefined as
\begin{align} \label{eq:input_X_partial}
X&=\left(\mathbf{f}_m(t_{k}^{[\text{arr}]}), q_{m}(t_{k}^{[\text{arr}]}), v_k\alpha_{k},v_k\beta_{k}, r_k[m,n_m^*]\right).
\end{align}
Then, the proposed online task offloading policy and training methodology for minimizing the delay are similarly given as those in Sections \ref{subsec:online_policy} and \ref{subsec:q_learning}. The key difference is in the construction of state, action, and reward used for the DRL agent.
Specifically, Algorithm \ref{alg:state_gener_partial} describes the construction of $\mathcal{X}_k$.
As explained in Section \ref{subsec:online_policy}, all input values in Line 1 is available at the DRL agent.
In Line 5, we apply the delay estimator given in Section \ref{subsec:delay+prediction} to estimate the delay of edge computing for each server $m \in \mathcal{O}_k$ and task offloading ratio $v \in \mathcal{V}$. After constructing  $\Omega_{k}$ from $\hat\Gamma_{k}^{[\text{MEC}]}$ in Line 10 to Line 14, we finally obtain $\mathcal{X}_k$ in Line 15. Here, the role of the DRL agent is to determine the offloading server and the offloading ratio for local and edge computing among the $L$ nearest servers in $\mathcal{O}_k$ for $k=1,2,\cdots$. Therefore, the action space of $J_k$ is given by $\mathcal{A}_{k}= \{ (m,v)| m \in \mathcal{O}_{k}, v \in \mathcal{V} \}$. The DRL agent takes the state $\mathcal{X}_k$ as an input and generate $M\times|\mathcal{V}|$ Q-values. Then the action of the DRL agent for $J_k$ is determined by the maximum of the estimated Q-values in $\mathcal{A}_k$. Finally, the reward of $J_k$ is defined as 
\begin{align} \label{eq:reward_partial}
\Lambda_{k}= -\max \left\{ \tau^{[\text{local}]}_k, \tau_k^{[\text{MEC}]} \right\}.
\end{align}
 
\begin{algorithm}[!t] 
\caption{State generation of the DRL agent for partial offloading.}\label{alg:state_gener_partial}
\begin{algorithmic}[1]
\State{{\bf Input}: $\mathbf{u}_k^{[\text{user}]}$, $\alpha_k$, $\beta_k$, $\{\mathbf{f}_m(t_{k}^{[\text{arr}]})\}_{m\in \mathcal{O}_k}$, $\{q_m(t_k^{[\text{arr}]})\}_{m\in \mathcal{O}_k}$, $\{r_k[m,n_m^*]\}_{m\in \mathcal{O}_k}$, and $\mathbf{a}_{k}=[a_{k-1},a_{k-2}, \cdots,a_{k-A}]$.}
\State{{\bf Initialization}: Set $\hat\Gamma_{k}^{[\text{MEC}]}=\mathbf{0}_{M\times|\mathcal{V}|}$, $  \Omega_{k}=\mathbf{0}_{M\times|\mathcal{V}|}$, 
 and $\mathcal{A}= \{ (m,v)| m \in \mathcal{O}_{k}, v \in \mathcal{V} \}$.}
 
\For{all $m \in \mathcal{O}_k$}
\For{all $v \in \mathcal{V}$}
\SplitState{Set $\big( \mathbf{f}_m(t_{k}^{[\text{arr}]}), q_{m}(t_{k}^{[\text{arr}]}), v\alpha_{k}, v\beta_{k}, r_k[m,n_m^*]\big)$ as the input of the delay estimator and obtain the estimated delay  $\hat\tau_{k,m}^{[\text{MEC}]}$, which corresponds to the output value of the estimator.}
\State{Calculate the local delay $\tau_{k}^{[\text{local}]}$ using \eqref{eq:local_partial}.}
\State{Update $[\hat\Gamma_{k}^{[\text{MEC}]}]_{m,v} \leftarrow \max \{ \tau_{k}^{[\text{local}]},\hat\tau_{k,m}^{[\text{MEC}]} \}$.}
\EndFor
\EndFor
\For{$j \in [1:L]$}
\State{Set $(m',v')=\underset{(m,v) \in \mathcal{A}}{\arg \min}{[\hat\Gamma_{k}^{[\text{MEC}]}]_{m,v}}.$}
\State{Update $[\Omega_{k}]_{m',v'} \leftarrow j.$}
\State{Update $\mathcal{A} \leftarrow \mathcal{A} \setminus \{ (m',v') \}$.}
\EndFor
\State{{\bf Output}: $\mathcal{X}_k = (\Omega_k, \mathbf{a}_{k}$).}
 \end{algorithmic}
\end{algorithm}

\section{Numerical Evaluation} \label{sec:simulation}
In this section, we evaluate the performance of the proposed scheme and compare it with five benchmark schemes. For the first scheme, i.e., `Probabilistic offloading', the offloading decision for each task is taken with probability $p/L$ among $L$ nearest servers, where $p\in[0,1]$ is numerically optimized to minimize the average cost in simulation. Specifically, we repeated simulations for different values of $p$, ranging from $0$ to $1$ with a step size of $0.1$, and selected the value of $p$ 
 resulting in the minimum average cost. For the second and third schemes, i.e., `MEC only' and `Local only' respectively, all tasks will either be offloaded to the nearest MEC server for computation or computed locally. For the fourth scheme, i.e., `DRL benchmark', we consider the conventional DRL-based approach. The previous works in \cite{Ming_DRLoffloadqueue,chenyingDTODTMEC
,alfakihTORAMECDRL} utilize $(q_{m}(t_{k}^{[\text{arr}]}), \alpha_{k}, \beta_{k},\mathbf{a}_{k})$ for offloading decision and, for fair comparison,  we define the state of the DRL agent as $\mathcal{X}_k=\big(\mathbf{f}_m(t_{k}^{[\text{arr}]}), q_{m}(t_{k}^{[\text{arr}]}), \alpha_{k}, \beta_{k}, r_k[m,n_m^*], \mathbf{a}_{k}\big)$, which is the same set of information used for the proposed scheme.
 

We further consider a cost lower bound, denoted by `Lower bound'.  For the case, we assume the ideal case in which global and non-causal information of the entire MEC system is known and the offloading decision for each task is taken based on such information. That is, the agent can calculate the exact delay in advance and optimally assign each task either to its local device or one of the $L$ nearest MEC servers.

Table \ref{tab:table1} summarizes the main system parameters and their values used in simulation, which was performed on a desktop computer with an AMD Ryzen 7 3800X 8-Core 3.89 GHz processor, NVIDIA GeForce RTX 2070 SUPER graphics card, and 32 GB memory. The parameter values for the task data size, task computation size, local computation capability, and MEC computation capability were set to similar values used in \cite{DaixingxiaTCOD2DMEC,Ming_DRLoffloadqueue}. We employ two fully connected DNNs for predicting the delay and energy consumption, respectively, each consisting of three hidden layers with $1000$, $2000$, and $2000$ neurons. The delay and energy estimators are trained in an offline manner based on supervised learning with $10^5$ data samples, which can be attained after completing the computing of each task, and then the same trained DNN is used throughout the entire simulation episode.
Recall that double Q-learning with $\mbox{DQN}^{\alpha}$ and $\mbox{DQN}^{\beta}$ are used for the DRL agent. In simulation, both $\mbox{DQN}^{\alpha}$ and $\mbox{DQN}^{\beta}$ are set to have two hidden layers with $512$ neurons in each layer and are trained in an online manner based on real-time experience.
That is, $\mbox{DQN}^{\alpha}$ and $\mbox{DQN}^{\beta}$ are continuously trained during the simulation episode, which allows the DRL agent to adapt its offloading decisions according to environmental change.

\subsection{Average Delay Minimization}

In this subsection, we firstly evaluate the performance of the proposed and benchmark schemes focused on the average delay. Note that by setting  $\omega^{[\text{d}]}=1$, $\omega^{[\text{e}]}=0$, and $C_k^{[{\text d}]}= \tau_k$ in \eqref{eq:cost}, the general cost function in Section \ref{subsec:general_cost} becomes the delay metric considered in Section \ref{subsec:delay_min}.

\begin{figure}[]
\centering
     \begin{subfigure}[]{0.49\textwidth}
\centering
         \includegraphics[width=0.9\linewidth]{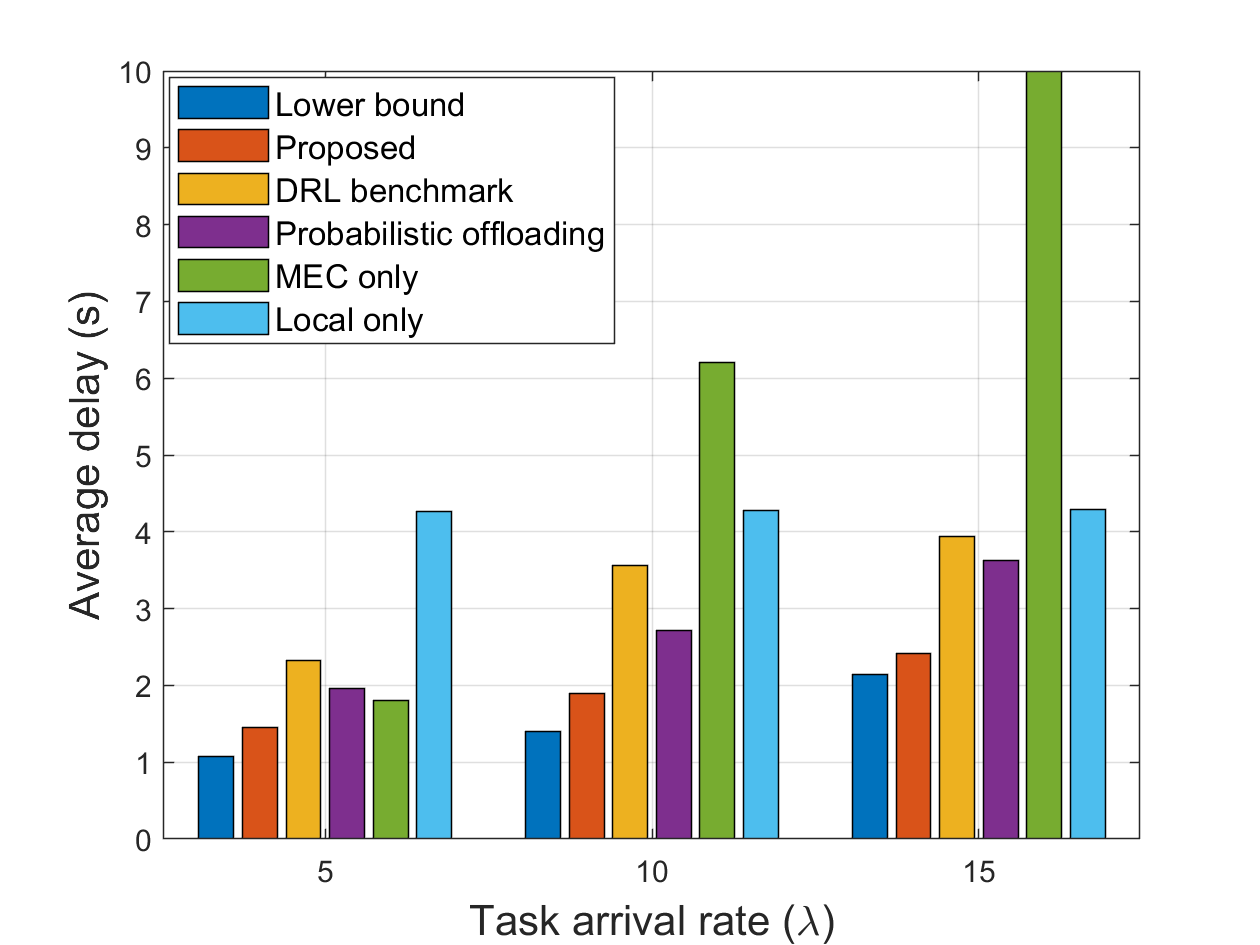}
         \caption{Average delay with respect to $\lambda$}
     \end{subfigure}
\hfill
     \begin{subfigure}[]{0.49\textwidth}
\centering
         \includegraphics[width=0.9\linewidth]{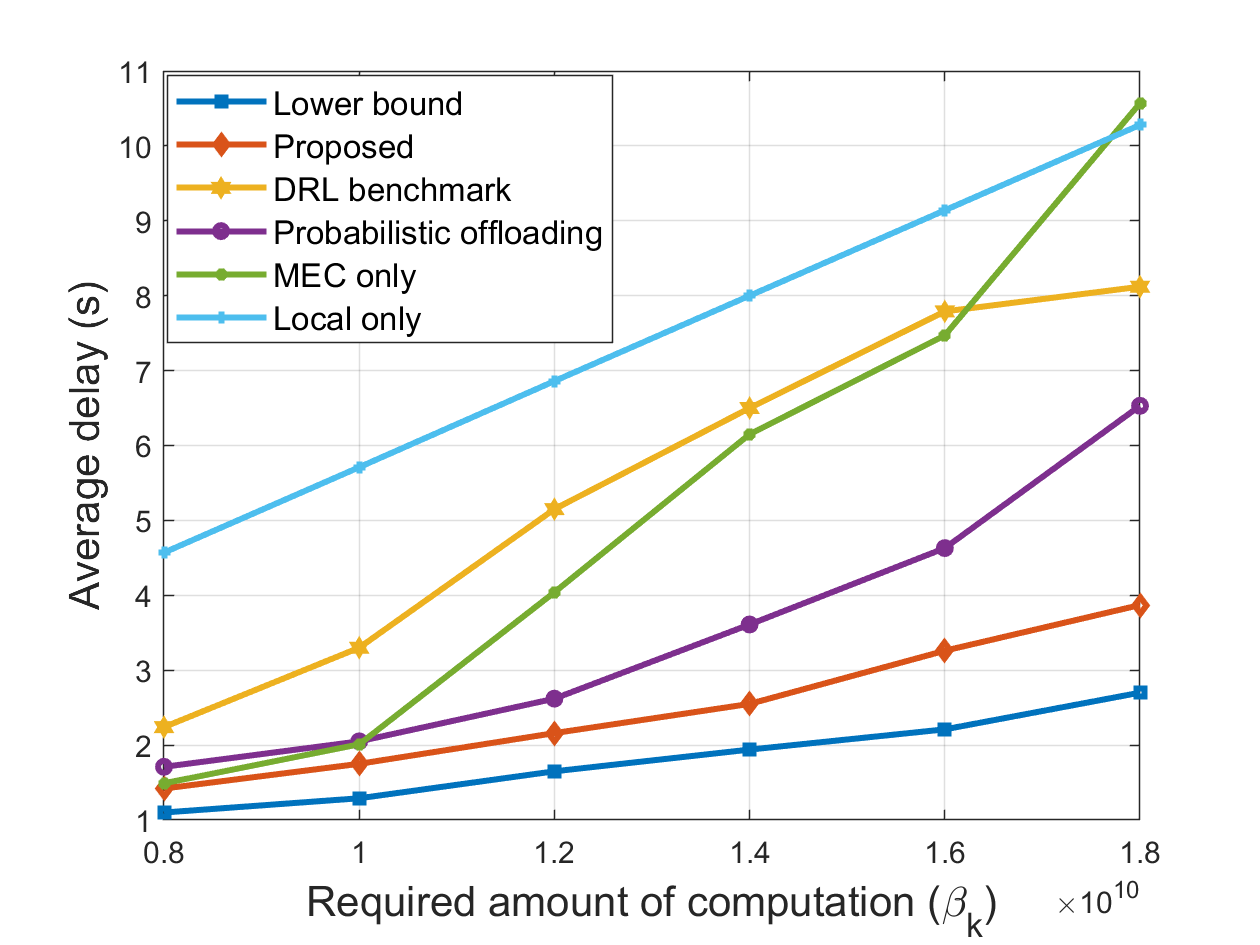}
         \caption{Average delay with respect to $\beta_k$}
     \end{subfigure}
        \caption{Average delay with respect to $\lambda$ and $\beta_k$.}
        \label{fig:varying_arrival_rate_compu_delay}
\end{figure}

Fig. \ref{fig:varying_arrival_rate_compu_delay} (a) plots the average delays of the proposed and three benchmark schemes with respect to $\lambda$, where $\lambda$ denotes the task arrival rate. As seen in the figure, the proposed scheme achieves universally good performance and strictly outperforms these four benchmark schemes.
On the other hand, there exist performance trade-offs between `Probabilistic offloading', `MEC only', and `Local only' depending on the task arrival rate $\lambda$. In particular, as $\lambda$ increases, `Local only' will be eventually better than `MEC only' due to excessive queuing delay of edge computing. More importantly, even though the fraction of edge and local computing is numerically optimized for `Probabilistic offloading', the proposed scheme can further reduce the overall delay by accurately estimating the expected delay for edge computing of each candidate server. 
For instance, the proposed scheme significantly decreases the overall delay by $29\%$, $78\%$, $49\%$, and $39\%$ compared to the cases of `Probabilistic offloading', `MEC only', `Local only', and `DRL benchmark', respectively when $\lambda=15$.
Fig. \ref{fig:varying_arrival_rate_compu_delay} (b) plots the average delays with respect to the required amount of computation $\beta_k$, where we assume $\beta_k$ is the same as $\beta$ for all $k$. As seen in the figure, the proposed scheme again outperforms all the benchmark schemes for all possible ranges of $\beta$.

\begin{figure}[]
\centering
     \begin{subfigure}[]{0.49\textwidth}
\centering
         \includegraphics[width=0.9\linewidth]{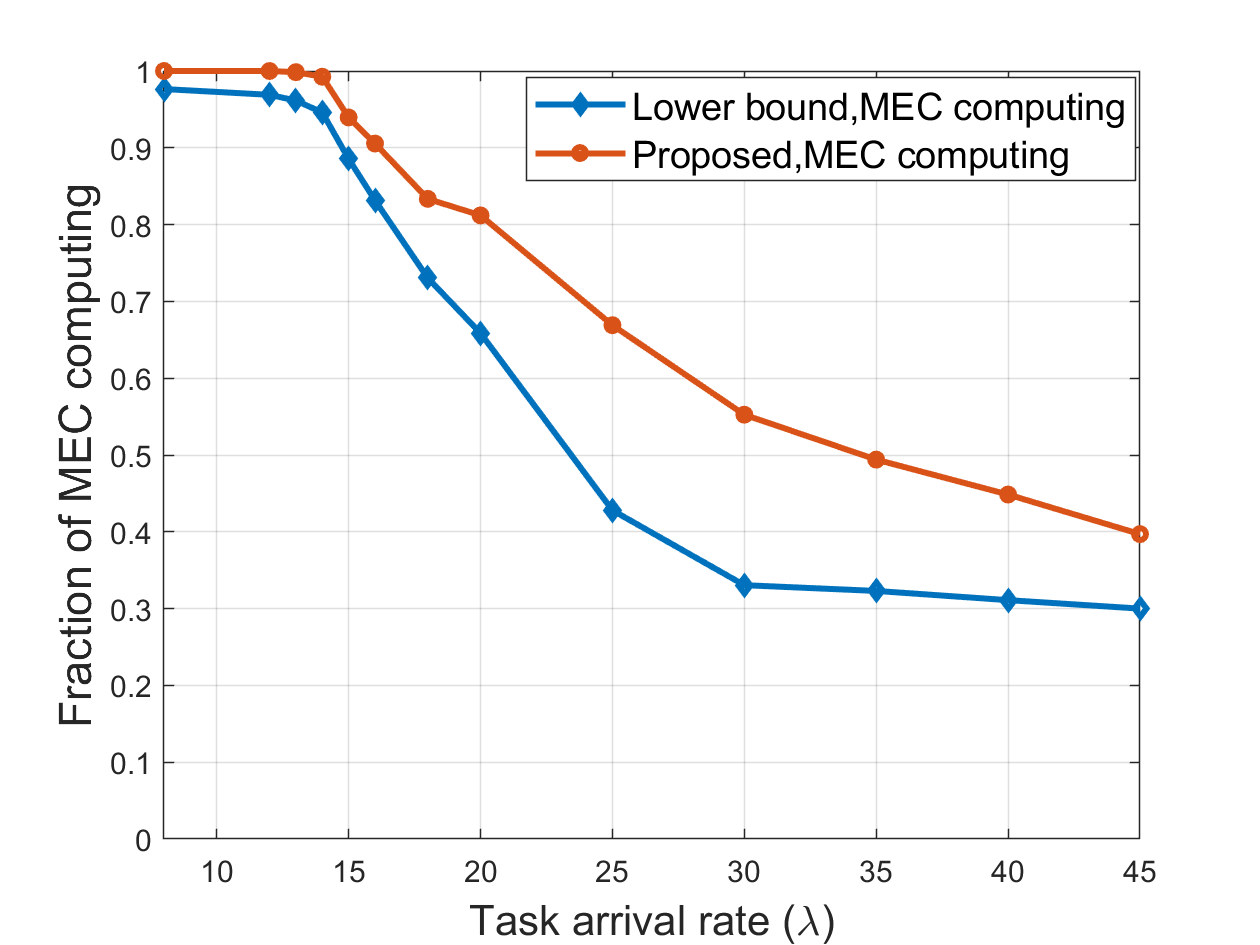}
         \caption{Fraction of MEC computing with respect to the task arrival rate}
     \end{subfigure}
\hfill
     \begin{subfigure}[]{0.49\textwidth}
\centering
         \includegraphics[width=0.9\linewidth]{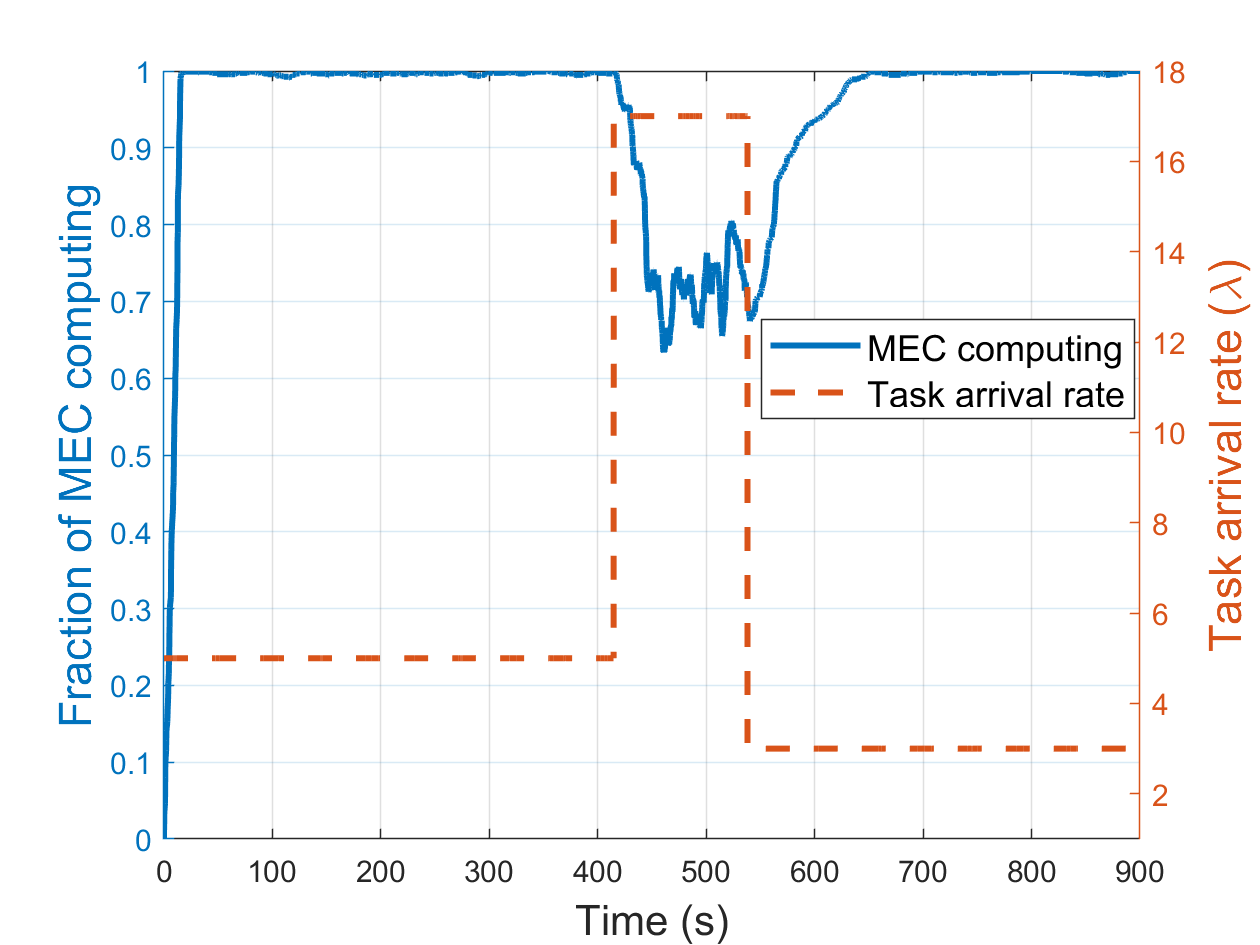}
         \caption{Fraction of MEC computing with respect to time varying task arrival rate}
     \end{subfigure}
        \caption{Fraction of MEC computing with respect to $\lambda$.}
        \label{fig:vary_lam_local_mec_ratio}
\end{figure}

Fig. \ref{fig:vary_lam_local_mec_ratio} (a) plots the fraction of MEC computing with respect to the task arrival rate $\lambda$. As $\lambda$ increases, offloading tasks to edge servers results in large queuing delay, therefore the fraction of local computing increases. The proposed scheme can efficiently adjust its task offloading strategy according to such network dynamics. Fig. \ref{fig:vary_lam_local_mec_ratio} (b) plots the fraction of MEC computing of the proposed scheme with respect to the time varying task arrival rate $\lambda$. As $\lambda$ varies over time, the proposed scheme is able to adapt its task offloading strategy in real-time. The figure shows that more MEC computing occurs with lower values of $\lambda$, and more local computing occurs with higher values of $\lambda$ to avoid large queuing delay.

\begin{figure}[]
\centering
     \begin{subfigure}[]{0.49\textwidth}
\centering
         \includegraphics[width=0.9\linewidth]{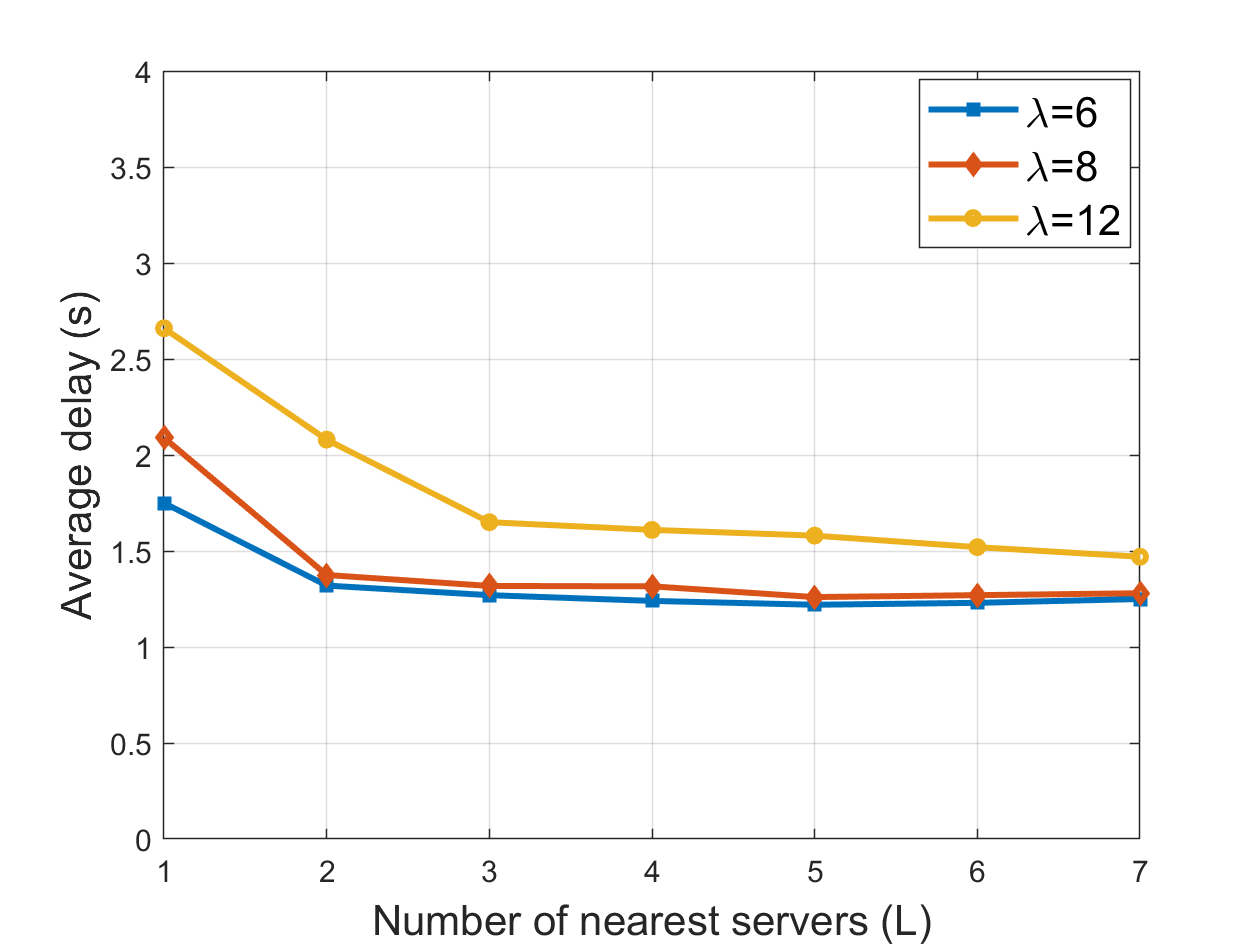}
         \caption{Average delay with respect to varying $\lambda$}
     \end{subfigure}
\hfill
     \begin{subfigure}[]{0.49\textwidth}
\centering
         \includegraphics[width=0.9\linewidth]{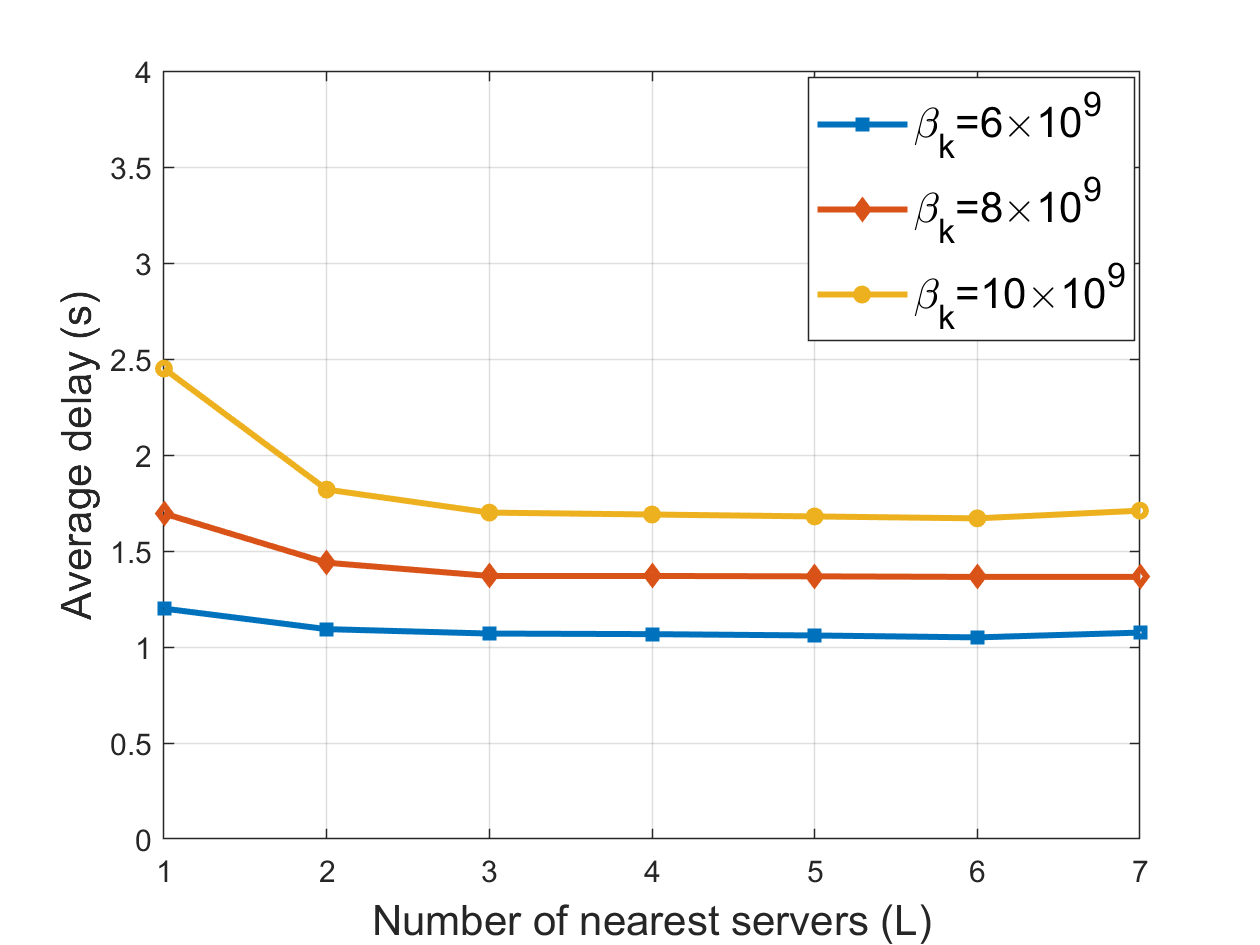}
         \caption{Average delay with respect to varying $\beta_k$}
     \end{subfigure}
        \caption{Average delay with respect to the number of nearest servers.}
        \label{fig:varying_nearby_server_arrivalrates_final}
\end{figure}

Fig. \ref{fig:varying_nearby_server_arrivalrates_final} (a) and Fig. \ref{fig:varying_nearby_server_arrivalrates_final} (b) plot the average delay of the proposed scheme with respect to the nearest servers $L$, which is the number of candidate edge servers for MEC. As seen in the figures, $L=3$ is enough to stabilize the average delay performance. 
Therefore, the proposed scheme can reduce signaling and computation overhead by $L/M=1/5$ for this case while preserving its delay performance compared to the case where global network information is available. As a consequence, the proposed edge computing framework and learning methodology provide a scalable solution applicable even as the network size increases.

\subsection{General Cost Functions}
\begin{figure}[]
\centering
     \begin{subfigure}[]{0.49\textwidth}
\centering
         \includegraphics[width=0.9\linewidth]{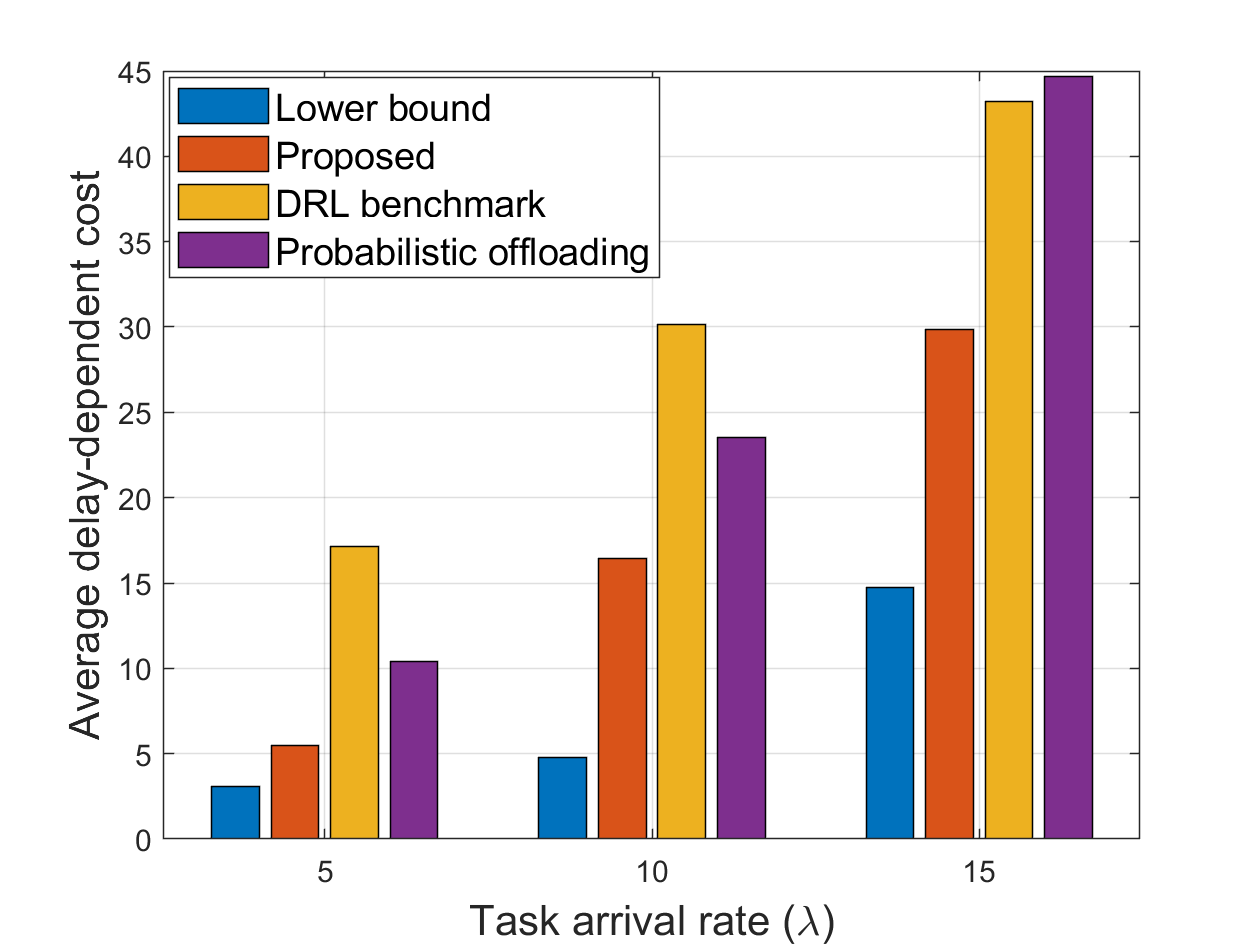}
         \caption{Exponential cost function}
     \end{subfigure}
\hfill
\begin{subfigure}[]{0.49\textwidth}
\centering
         \includegraphics[width=0.9\linewidth]{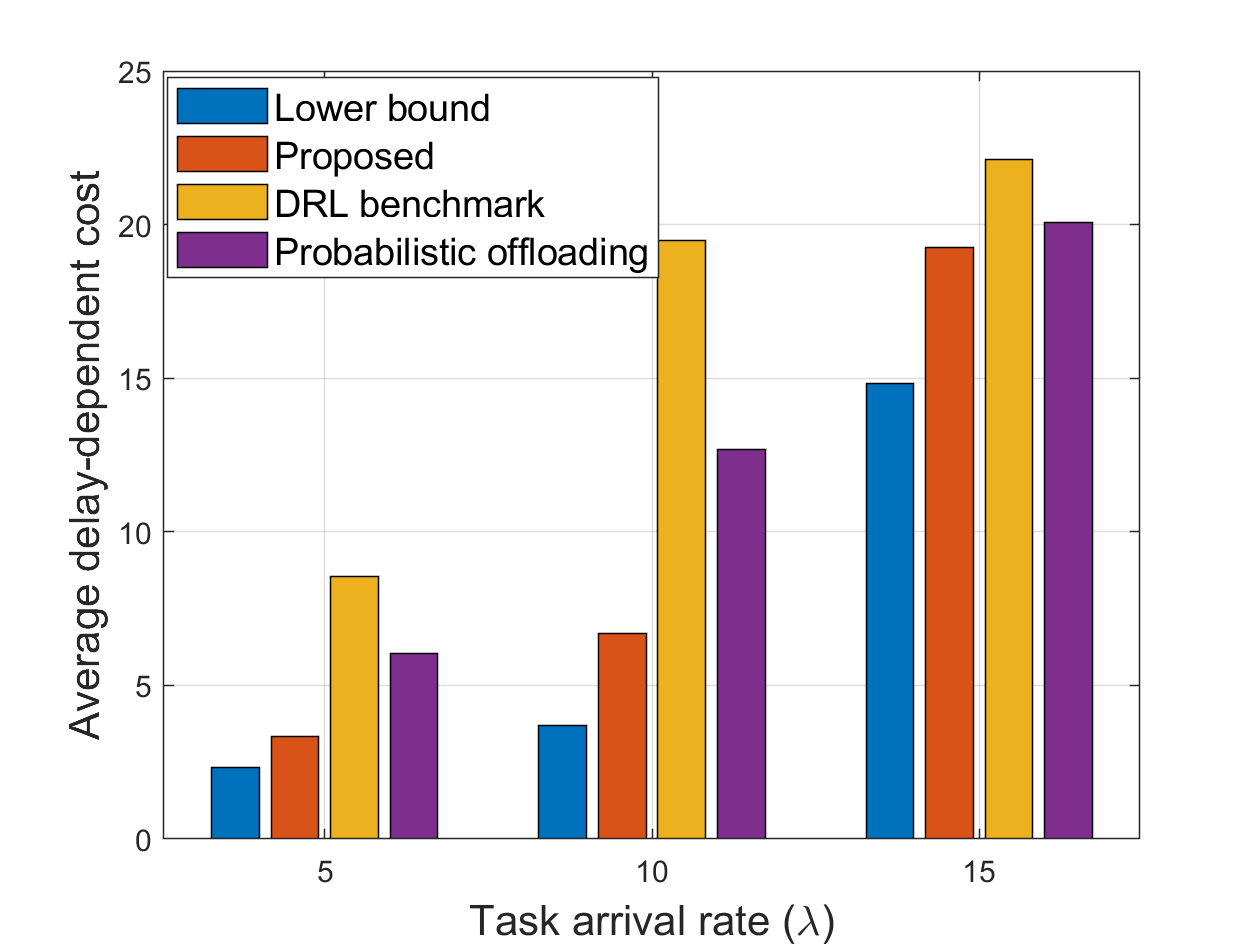}
         \caption{Quadratic cost function}
     \end{subfigure}

        \caption{Average delay-dependent cost with respect to $\lambda$ for different cost functions.}
        \label{fig:varying_arrival_rate_cost_delay}
\end{figure}

In this subsection, we consider general cost functions in Section \ref{subsec:general_cost}.
In order to demonstrate that the proposed scheme can achieve universally good performance irrespective of the types of delay-dependent cost functions, we evaluate the performance of the propose scheme with three different delay-dependent cost functions, i.e., exponential, quadratic, and logarithmic functions of the delay. Fig. \ref{fig:varying_arrival_rate_cost_delay} plots the performance of the proposed scheme with respect to exponential and quadratic delay-dependent cost functions, respectively, which might be suitable for delay-sensitive applications.
As seen in the figures, the proposed scheme provides universally improved performance for all cost functions compared to `Probabilistic offloading', and `DRL benchmark' task offloading schemes.

\begin{figure}[]
\centering
     \begin{subfigure}[]{0.49\textwidth}
\centering
         \includegraphics[width=0.9\linewidth]{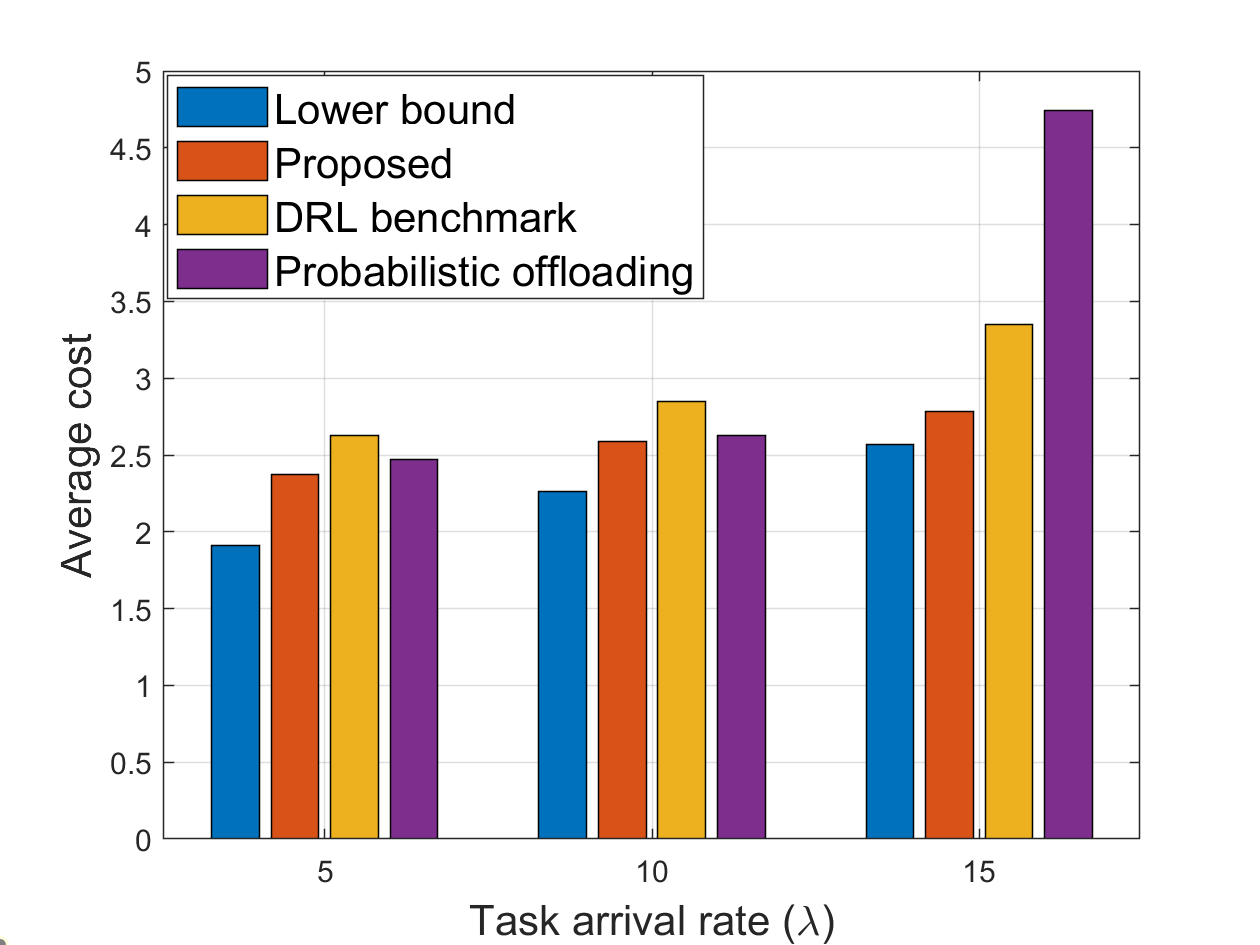}
         \caption{Average cost with respect to $\lambda$}
     \end{subfigure}
\hfill
     \begin{subfigure}[]{0.49\textwidth}
\centering
         \includegraphics[width=0.9\linewidth]{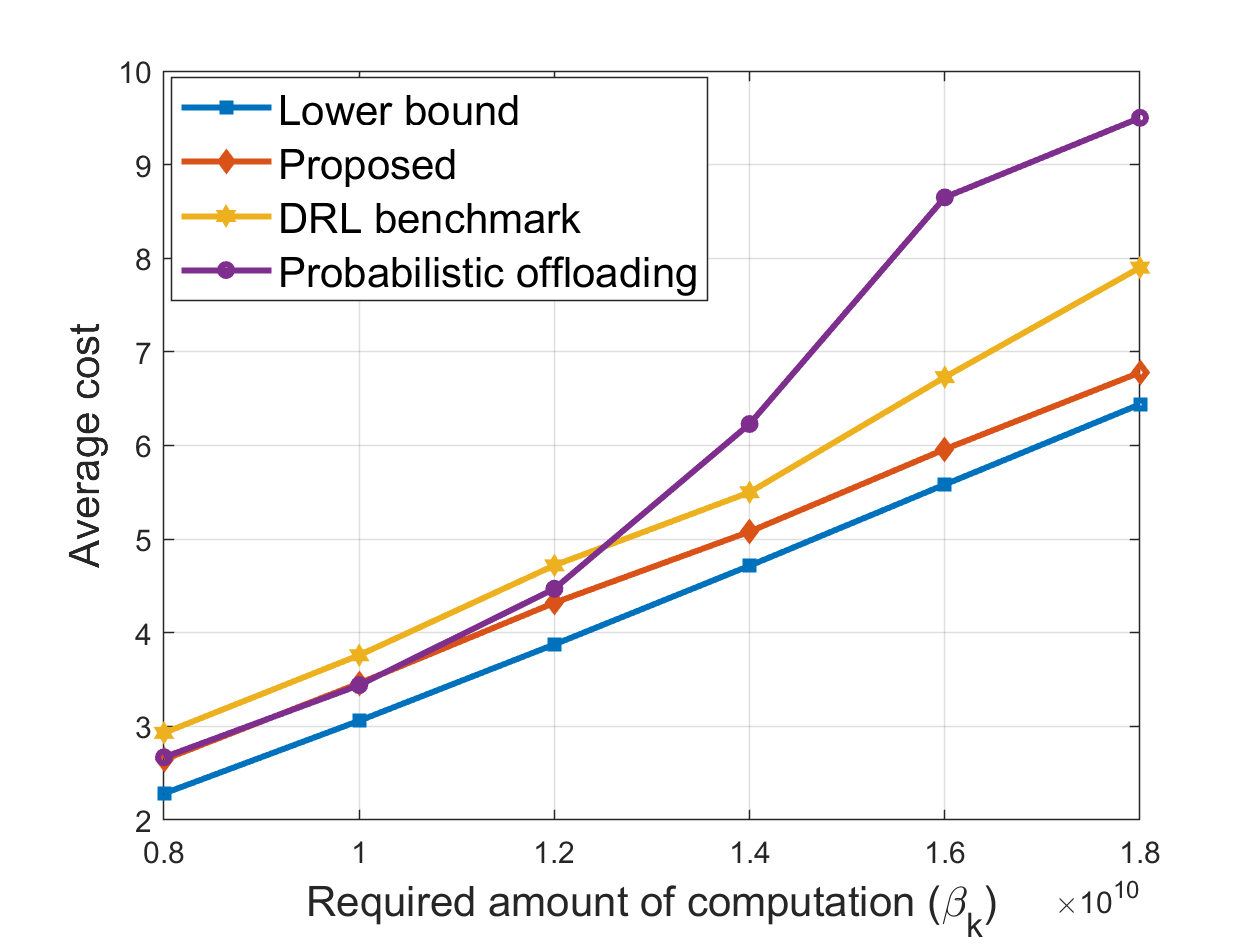}
         \caption{Average cost with respect to $\beta_k$}
     \end{subfigure}
        \caption{Average cost with respect to $\lambda$ and $\beta_k$.}
        \label{fig:varying_arrival_rate_compu_delay_energy_cost}
\end{figure}

Fig. \ref{fig:varying_arrival_rate_compu_delay_energy_cost} (a) plots the average cost reflecting both energy and delay metrics, i.e., $C^{[\text{d}]}_k$ and $C^{[\text{e}]}_k$ in \eqref{eq:cost} with $w^{[\text{d}]}=0.5$ and $w^{[\text{e}]}=0.5$, with respect to $\lambda$. As seen in the figure, the proposed scheme is able to obtain improved performance comparable to the benchmark scheme.
Similarly, Fig. \ref{fig:varying_arrival_rate_compu_delay_energy_cost} (b) plots the average cost by setting $w^{[\text{d}]}=0.5$ and $w^{[\text{e}]}=0.5$, with respect to the required amount of computation $\beta_k$. As $\beta_k$ increases, both delay- and energy-dependent costs  rapidly increases for the case of probabilistic offloading, demonstrating that appropriate selection of an edge server by accurately estimating the expected delay and consumed energy is crucially important. 
As seen in the figure, the average cost of the proposed scheme maintains within a constant gap from the cost lower bound and performs better than the `DRL benchmark' scheme.

\subsection{Partial Offloading}
\begin{figure}
  \centering
  \begin{subfigure}{0.8\columnwidth}
    \centering
    \includegraphics[width=\textwidth]{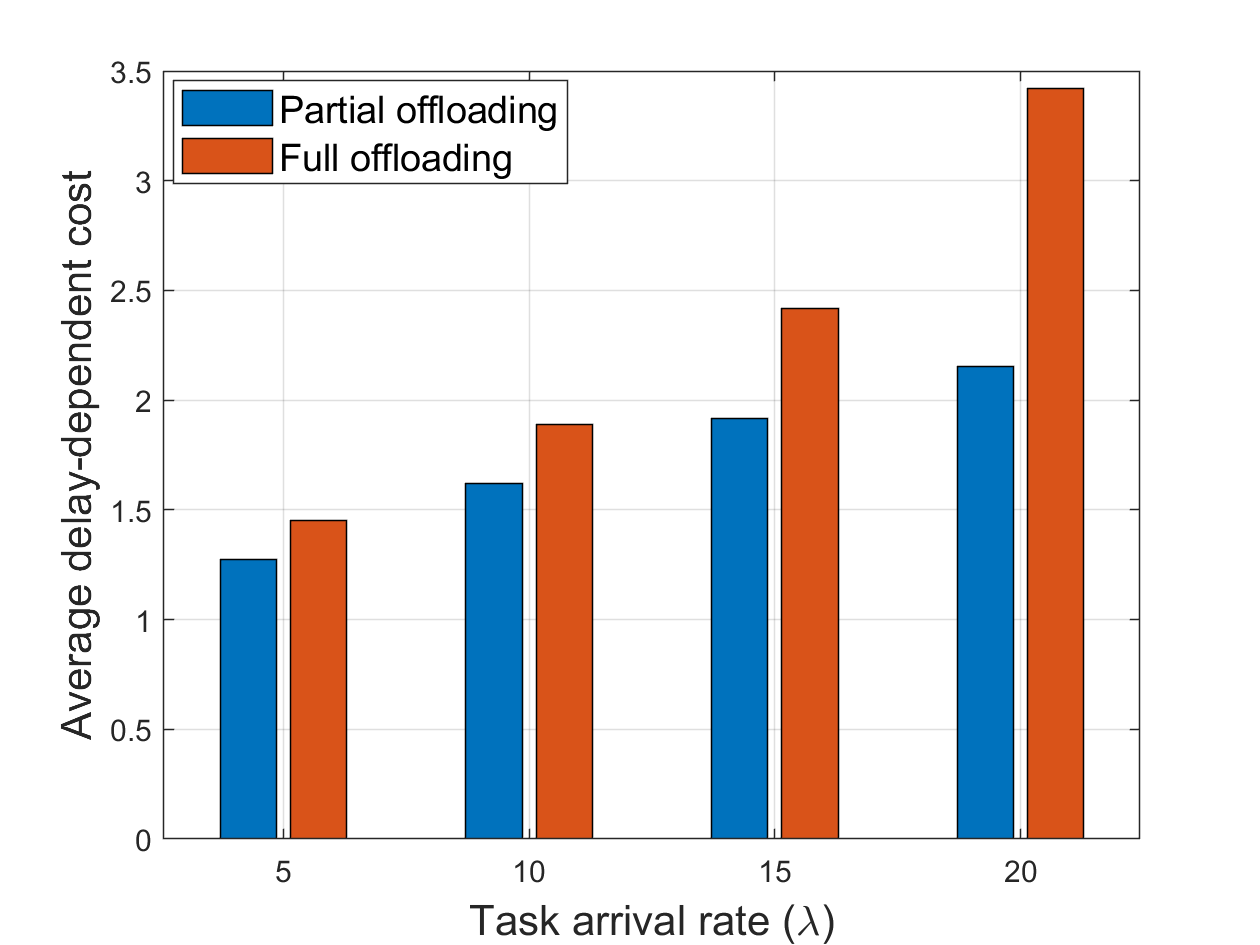}
    \caption{Linear cost function}
  \end{subfigure}%
  \hspace{0.05\columnwidth} 
  \begin{subfigure}{0.8\columnwidth}
    \centering
    \includegraphics[width=\textwidth]{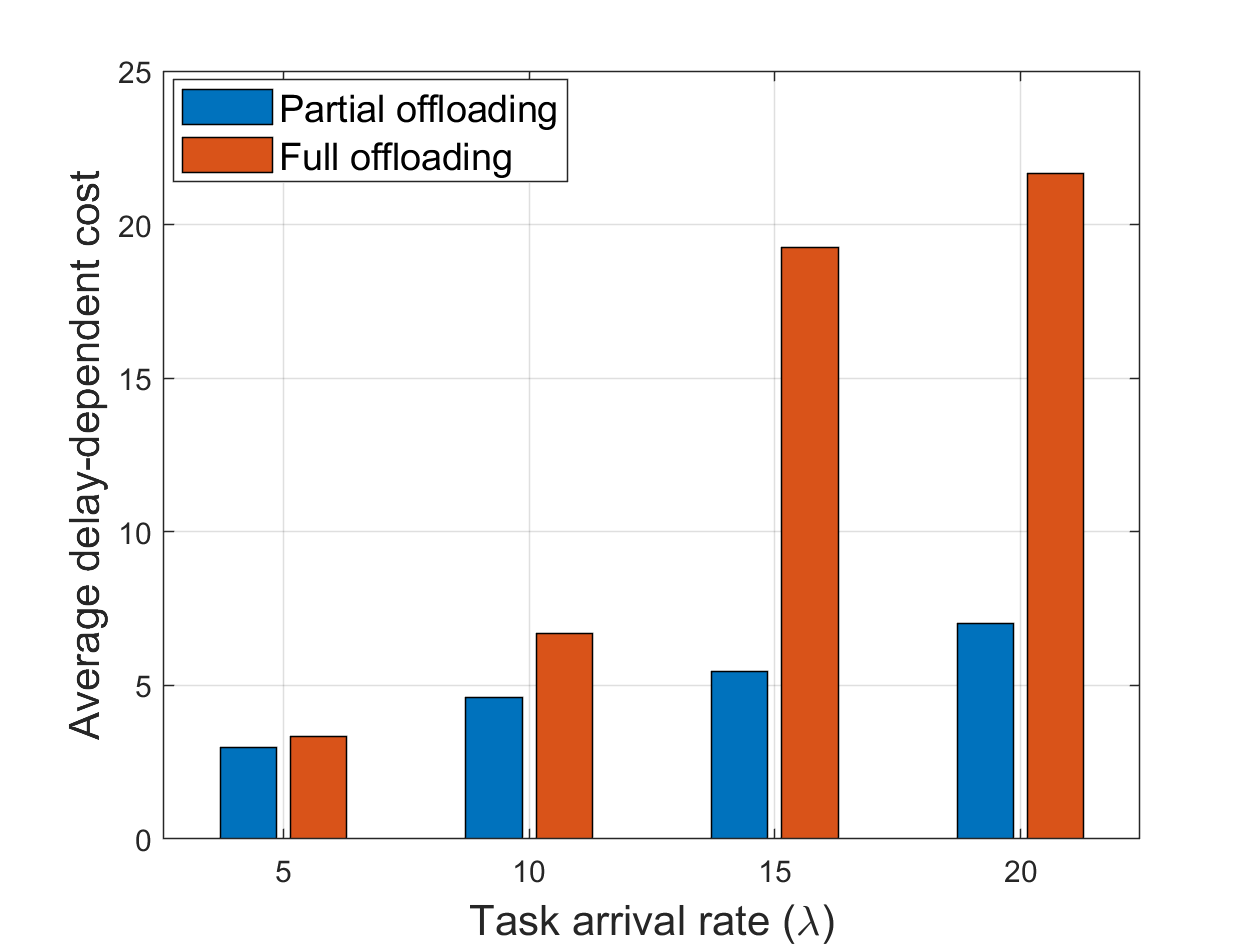}
    \caption{Quadratic cost function}
  \end{subfigure}
  
  \vspace{0.05\columnwidth} 
  
  \begin{subfigure}{0.8\columnwidth}
    \centering
    \includegraphics[width=\textwidth]{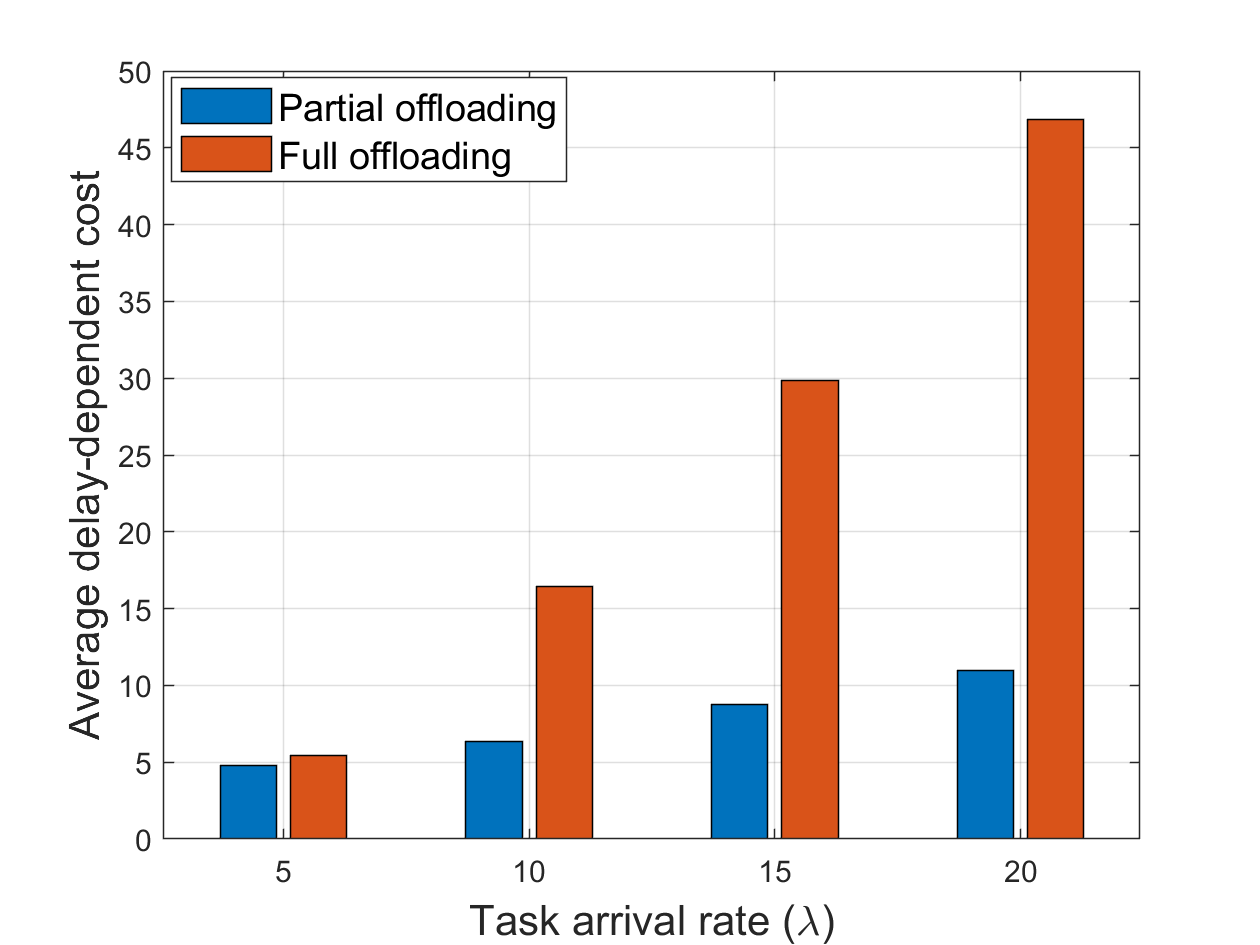}
    \caption{Exponential cost function}
  \end{subfigure}
  
  \caption{Average delay-dependent cost with respect to $\lambda$ for
different cost functions.}
\label{fig:varying_arrival_rate_cost_delay_partial}
\end{figure}

In this subsection, we consider partial offloading stated in Section \ref{subsec:partial_offload}. In order to demonstrate the impact of partial offloading, we evaluate the performance of the proposed partial offloading scheme and compare it with the full offloading case under three different delay-dependent cost functions, i.e., linear, quadratic, and exponential functions of delay. We assume the same simulation parameters used in Fig. \ref{fig:varying_arrival_rate_cost_delay} and set $\mathcal{V}=[0.1,0.2,\cdots,0.9]$. Fig. \ref{fig:varying_arrival_rate_cost_delay_partial} plots the performance of the proposed partial offloading scheme with respect to the task arrival rate $\lambda$ and  Table \ref{tab:table2} illustrates the percentage improvement achieved by the proposed partial offloading scheme in reducing the average delay-dependent cost compared to the full offloading case. The results reveals that partial offloading proves particularly advantageous for delay-sensitive applications, leveraging parallel computing capabilities at both the local device and edge server.

\begin{table}[h]
\centering
\caption{Percentage improvement in average delay-dependent cost reduction.}
\label{tab:table2}
\scalebox{1.1}{%
\begin{tabular}{c|c|c|c|c}
\hline

Cost function & \(\lambda = 5\) & \(\lambda = 10\) & \(\lambda = 15\) & \(\lambda = 20\) \\
\hline
\hline
Linear & 12.17 & 14.32 & 20.70 & 37.05 \\
\hline
Quadratic & 11.52 & 31.28 & 71.68 & 67.58 \\
\hline
Exponential & 11.83 & 61.34 & 73 & 76.54 \\
\hline
\end{tabular}%
}
\end{table}

\section{Conclusion}\label{sec:conclusion}
In this paper, we proposed a novel hybrid online--offline learning based task offloading policy that can efficiently reduce the average computation delays of MEC systems by adjusting its offloading policy depending on network dynamics with limited local network information. Numerical results demonstrated that the proposed offloading scheme outperforms the conventional MEC schemes and it is universally applicable for a broad classes of network configurations. 
It has been further shown that the proposed hybrid online--offline learning framework can be extended to a general cost function reflecting both delay- and energy-dependent metrics. 


\section*{Appendix \\ Derivation of queuing and computation delay}
In this appendix, we state the calculation of $\tau^{[\text{queue}]}_k$ and $\tau^{[\text{comp}]}_k$ in \eqref{eq:delay_definition} and \eqref{eq:_delay_edge}, which are required to evaluate the performance of the proposed and benchmark schemes. For the proposed scheme, \eqref{eq:delay_definition} and \eqref{eq:_delay_edge} are also required to calculate the reward in \eqref{eq:reward}.
Let us define the set of tasks $\mathcal{J}_m(t)$ consisting of all assigned tasks to server $m$ that have been completely uploaded to server $m$ before time $t$ but their computation has not started at server $m$ until time $t$. That is, $\mathcal{J}_m(t)$ is the set of tasks stored in the queue memory of server $m$ at time $t$.
From the definition of $\mathcal{J}_m(t)$, we have $\mathcal{J}_m(t)=\left\{J_l| s_l[m]=1 \text{ for all } l \text{ such that } t^{[\text{queue}]}_l\leq t \leq t^{[\text{comp}]}_l\right\}$.
Note that $\mathcal{J}_m(t)$ can be the empty set. If $\mathcal{J}_m(t)= \emptyset$, then there can exist either a task which is being computed at time $t$ or not.
On the other hand, there always exists a task which is being computed at time $t$ if $\mathcal{J}_m(t)\neq \emptyset$. Denote such task as $J_{k^*_m(t)}$ that satisfies $t^{[\text{comp}]}_{k^*_m(t)}\leq t \leq t^{[\text{dep}]}_{k^*_m(t)}$.
We further define $q_m(t)$ as the amount of remaining CPU cycles required to complete the tasks in $\mathcal{J}_m(t)\cup\{J_{k^*_m(t)}\}$ at time $t$.
\begin{align}\label{eq:queue_siz}
q_m(t)=\sum_{l\in \mathcal{J}_m(t)} \beta_l+\beta_{k^*_m(t)}-\bar{\beta}_{k^*_m(t)},
\end{align}
where $\bar{\beta}_{k^*_m(t)}$ denotes the amount of CPU cycles for $J_{k^*_m(t)}$ that have already been computed at time $t$.
To specify $\bar{\beta}_{k^*_m(t)}$, define $i'$ such that $t^{[\text{up}]}_m[i']\leq t^{[\text{comp}]}_{k^*_m(t)}\leq t^{[\text{up}]}_m[i'+1]$. Similarly, define $i''$ such that $t^{[\text{up}]}_m[i'']\leq t\leq t^{[\text{up}]}_m[i''+1]$. Then, we have
\begin{align} \label{eq:beta_bar}
\bar{\beta}_{k^*_m(t)}=\begin{cases}
f_{m,i'}\left(t-t^{[\text{comp}]}_{k^*_m(t)}\right) &\mbox{ if } i''=i,\\
f_{m,i'}\left(t^{[\text{up}]}_m[i'+1]-t^{[\text{comp}]}_{k^*_m(t)}\right)\\
+\sum_{i=i'+1}^{i''-1}f_{m,i}\left(t^{[\text{up}]}_m[i+1]-t^{[\text{up}]}_m[i]\right)\\
+f_{m,i''}\left(t-t^{[\text{up}]}_m[i'+1]\right) &\mbox{ otherwise.}
\end{cases}
\end{align}
Note that if $i''=i'+1$, the term $\sum_{i=i'+1}^{i''-1}f_{m,i}\left(t^{[\text{up}]}_m[i+1]-t^{[\text{up}]}_m[i]\right)$ in \eqref{eq:beta_bar} becomes zero.

Recall that the task $J_k$ will be uploaded at time $t_k^{[\text{queue}]}$, $\tau_k^{[\text{queue}]}$ is determined by $q_m(t_k^{[\text{queue}]})$.
We define $j'$ such that $t_m^{[\text{up}]}[j']\leq t_k^{[\text{queue}]}\leq t_m^{[\text{up}]}[j'+1]$.
Let us first consider the case where $f_{m,j'}(t_m^{[\text{up}]}[j'+1]-t_k^{[\text{queue}]})>q_m(t_k^{[\text{queue}]})$.
For this case, 
\begin{align} \label{eq:tau_queue_1}
\tau_k^{[\text{queue}]}=\frac{q_m(t_k^{[\text{queue}]})}{f_{m,j'}}.
\end{align}
Obviously, $\tau_k^{[\text{queue}]}=0$ if $q_m(t_k^{[\text{queue}]})=0$.
Otherwise, find a non-negative integer value $z'$ such that
\begin{multline}
f_{m,j'}(t_m^{[\text{up}]}[j'+1]-t_k^{[\text{queue}]})+\sum_{z=1}^{z'} f_{m,j'+z}(t_m^{[\text{up}]}[j'+z+1]\\-t_m^{[\text{up}]}[j'+z])\leq q_m(t_k^{[\text{queue}]})\\
\leq f_{m,j'}(t_m^{[\text{up}]}[j'+1]-t_k^{[\text{queue}]})+\sum_{z=1}^{z'+1} f_{m,j'+z}(t_m^{[\text{up}]}[j'+z+1]\\-t_m^{[\text{up}]}[j'+z]).
\end{multline}
Then 
\begin{align} \label{eq:tau_queue_2}
\tau_k^{[\text{queue}]}=t_m^{[\text{up}]}[j'+z'+1]-t_k^{[\text{queue}]}+\frac{q_m(t_k^{[\text{queue}]})-\bar{q}_m(t_k^{[\text{queue}]})}{f_{m,j+z'+1}}
\end{align}
where $\bar{q}_m(t_k^{[\text{queue}]})=f_{m,j'}(t_m^{[\text{up}]}[j'+1]-t_k^{[\text{queue}]}))+\sum_{z=1}^{z'} f_{m,j'+z}(t_m^{[\text{up}]}[j'+z+1]-t_m^{[\text{up}]}[j'+z])$.
Note that $\sum_{z=1}^{z'} f_{m,j'+z}(t_m^{[\text{up}]}[j'+z+1]-t_m^{[\text{up}]}[j'+z])=0$ if $z'=0$.

As the same manner, we can calculate $t_k^{[\text{dep}]}$, which is determined by $\beta_k$.
Define $j''$ such that $t_m^{[\text{up}]}[j'']\leq t_k^{[\text{comp}]}\leq t_m^{[\text{up}]}[j''+1]$.
Let us first consider the case where $f_{m,j''}(t_m^{[\text{up}]}[j''+1]-t_k^{[\text{comp}]})>\beta_k$. For this case, 
\begin{align} \label{eq:tau_comp_1}
\tau_k^{[\text{comp}]}=\frac{\beta_k}{f_{m,j''}}.
\end{align}
Otherwise, find a non-negative integer value $z''$ such that 
\begin{multline}\label{eq:comp_data_req_f}
f_{m,j''}(t_m^{[\text{up}]}[j''+1]-t_k^{[\text{comp}]})+\sum_{z=1}^{z''} f_{m,j''+z}(t_m^{[\text{up}]}[j''+z+1]-\\t_m^{[\text{up}]}[j''+z])\leq \beta_k
\leq f_{m,j''}(t_m^{[\text{up}]}[j''+1]-t_k^{[\text{comp}]})\\+\sum_{z=1}^{z''+1} f_{m,j''+z}(t_m^{[\text{up}]}[j''+z+1]-t_m^{[\text{up}]}[j''+z]).
\end{multline}
Then 
\begin{align}  \label{eq:tau_comp_2}
\tau_k^{[\text{comp}]}=t_m^{[\text{up}]}[j''+z''+1]-t_k^{[\text{comp}]}+\frac{\beta_k-\bar{\beta}_k}{f_{m,j+z''+1}}
\end{align}
where
\begin{multline} \label{eq:bar_beta}
\bar{\beta}_k=f_{m,j''}(t_m^{[\text{up}]}[j''+1]-t_k^{[\text{comp}]})\\+\sum_{z=1}^{z''} f_{m,j''+z}(t_m^{[\text{up}]}[j''+z+1]-t_m^{[\text{up}]}[j''+z])
\end{multline}
Note that $\sum_{z=1}^{z''} f_{m,j''+z}(t_m^{[\text{up}]}[j''+z+1]-t_m^{[\text{up}]}[j''+z])=0$ if $z''=0$.   
Finally, from \eqref{eq:tau_trans}, \eqref{eq:tau_queue_1}, \eqref{eq:tau_queue_2}, \eqref{eq:tau_comp_1}, and \eqref{eq:tau_comp_2}, we have $\tau_k^{[\text{MEC}]}=\tau^{[\text{trans}]}_k+\tau^{[\text{queue}]}_k+\tau^{[\text{comp}]}_k$ as in \eqref{eq:_delay_edge}.


\bibliographystyle{IEEEtran}
\bibliography{IEEEabrv,References}

\begin{IEEEbiography}[{\includegraphics[width=1in,height=1.25in,clip,keepaspectratio]{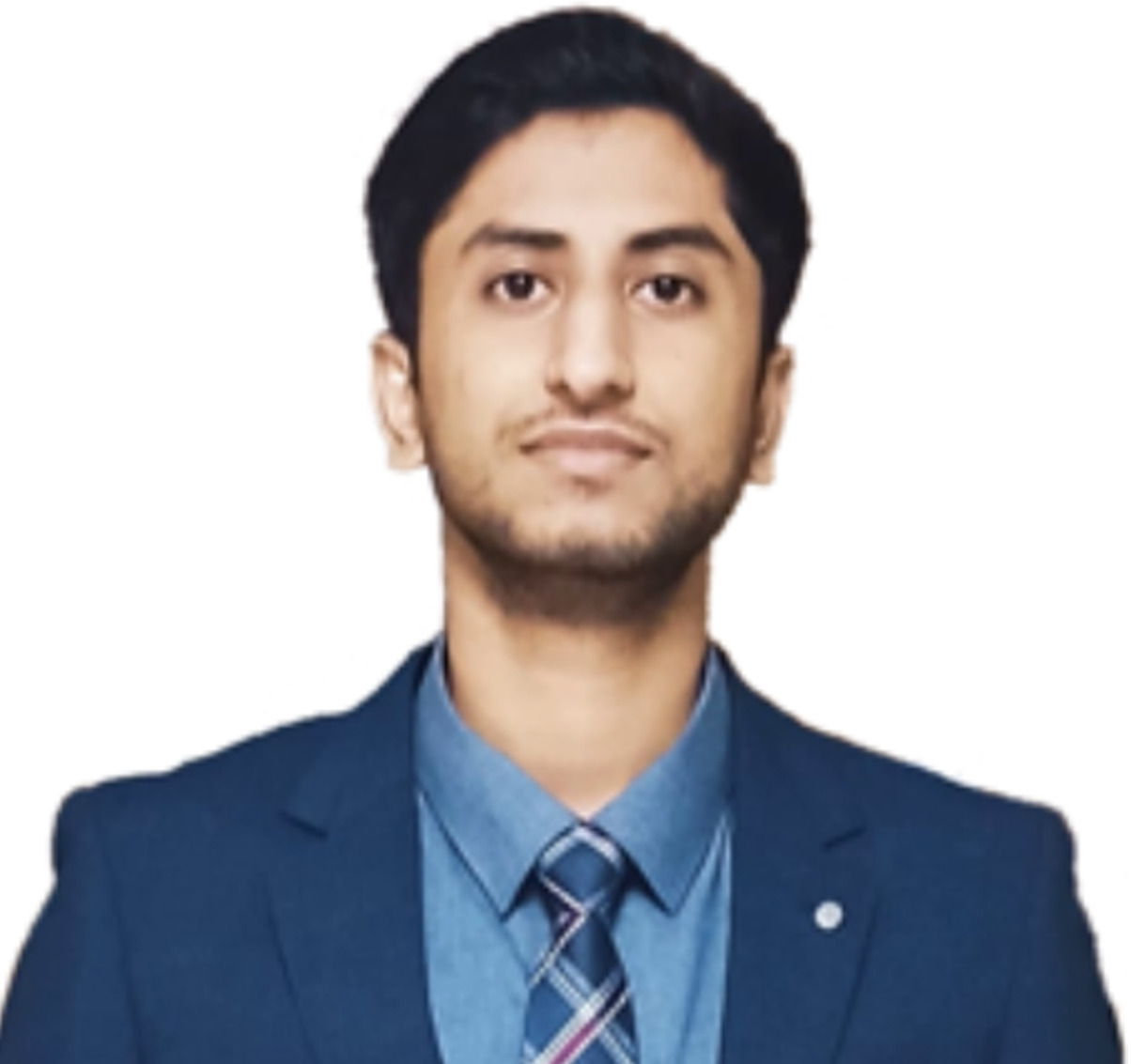}}]{Muhammmad Sohaib}
    received the B.S. degree in electrical engineering from the National University of Sciences and Technology (NUST), Islamabad, Pakistan, in 2018. He is currently pursuing the M.S. degree leading to the Ph.D. degree with Hanyang University, Ansan, South Korea. His research interests include reinforcement learning, random access, and mobile edge computing.
\end{IEEEbiography}
\begin{IEEEbiography}[{\includegraphics[width=1in,height=1.25in,clip,keepaspectratio]{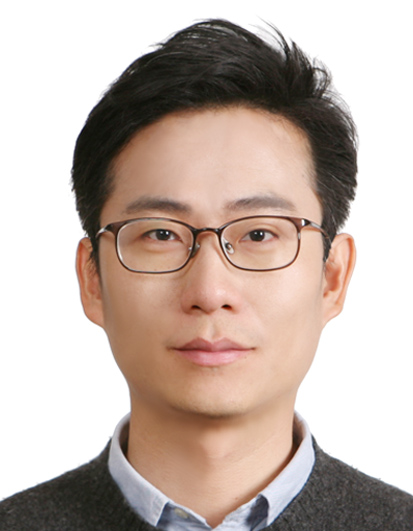}}]{Sang-Woon Jeon}
    (S'07--M'11--SM'23) received the B.S. and M.S. degrees in electrical engineering from Yonsei University, Seoul, South Korea, in 2003 and 2006, respectively, and the Ph.D. degree in electrical engineering from the Korea Advanced Institute of Science and Technology (KAIST), Daejeon, South Korea, in 2011. From 2011 to 2013, he was a Postdoctoral Associate at the School of Computer and Communication Sciences, Ecole Polytechnique Federale de Lausanne, Lausanne, Switzerland. . He is a Professor with the School of Computer Science and Technology, Zhejiang Normal University, Jinhua, China and the Department of Electrical and Electronic Engineering, Hanyang University, Ansan, South Korea. His research interests include wireless communication, evolutionary computation, and machine learning.
    He was a recipient of the Haedong Young Scholar Award in 2017, which was sponsored by the Haedong Foundation and given by the Korea Institute of Communications and Information Science (KICS), the Best Paper Award of the IEEE International Conference on Communications in 2015, and the Best Thesis Award from the Department of Electrical Engineering, KAIST, in 2012.
\end{IEEEbiography}
\begin{IEEEbiography}[{\includegraphics[width=1in,height=1.25in,clip,keepaspectratio]{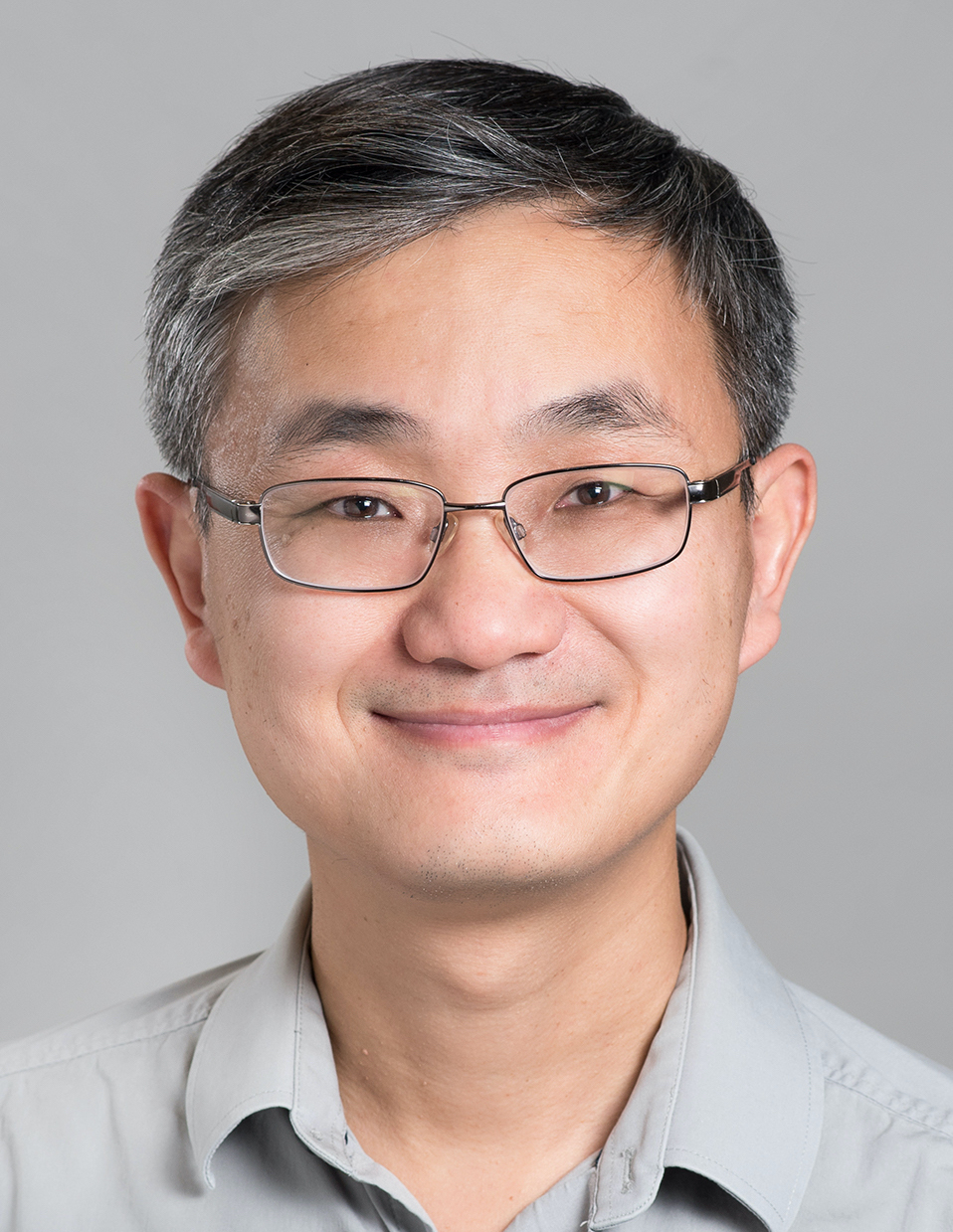}}]{Wei Yu}
    (Fellow, IEEE) received the B.A.Sc. degree in computer engineering and mathematics from the University of Waterloo, Waterloo, ON, Canada, and the M.S. and Ph.D. degrees in electrical engineering from Stanford University, Stanford, CA, USA. He is currently a Professor with the Department of Electrical and Computer Engineering, University of Toronto, Toronto, ON, Canada, where he holds a Canada Research Chair (Tier 1) in Information Theory and Wireless Communications. He is a Fellow of Canadian Academy of Engineering and a member of the College of New Scholars, Artists and Scientists, Royal Society of Canada. He received the Steacie Memorial Fellowship in 2015, the IEEE Marconi Prize Paper Award in Wireless Communications in 2019, the IEEE Communications Society Award for Advances in Communication in 2019, the IEEE Signal Processing Society Best Paper Award in 2008, 2017, and 2021, the Journal of Communications and Networks Best Paper Award in 2017, and the IEEE Communications Society Best Tutorial Paper Award in 2015. He was the President of the IEEE Information Theory Society in 2021. He served as the Chair for the Signal Processing for Communications and Networking Technical Committee of the IEEE Signal Processing Society from 2017 to 2018. He was an IEEE Communications Society Distinguished Lecturer from 2015 to 2016. He served as an Area Editor for IEEE {\sc Transactions on Wireless Communications}, an Associate Editor for IEEE {\sc Transactions on Information Theory}, and an Editor for  IEEE {\sc Transactions on Communications} and IEEE {\sc Transactions on  Wireless Communications}.
\end{IEEEbiography}

\end{document}